\documentclass[apj]{emulateapj}

\usepackage{natbib,amssymb,amsmath}
\usepackage{epstopdf}
\newcommand{\amm}{NH$_3$}
\newcommand{\dia}{N$_2$H$^+$}
\newcommand{\ddia}{N$_2$D$^+$}
\newcommand{\hd}{H$_2$D$^+$}
\newcommand{\co}{C$^{18}$O}

\newcommand{\kms}{km\,s$^{-1}$}
\newcommand{\cc}{cm$^{-3}$}

\newcommand{\showrev}{}

\newcommand{\noprint}[1]{}
\newcommand{\figsetstart}{{\bf Fig. Set} }
\newcommand{\figsetend}{}
\newcommand{\figsetgrpstart}{}
\newcommand{\figsetgrpend}{}
\newcommand{\figsetnum}[1]{{\bf #1.}}
\newcommand{\figsettitle}[1]{ {\bf #1} }
\newcommand{\figsetgrpnum}[1]{\noprint{#1}}
\newcommand{\figsetgrptitle}[1]{\noprint{#1}}
\newcommand{\figsetplot}[1]{\noprint{#1}}
\newcommand{\figsetgrpnote}[1]{\noprint{#1}}

\shorttitle{}
\shortauthors{Friesen et al.}

\begin{document}

\title{An analysis of the deuterium fractionation of star-forming cores in the Perseus molecular cloud}
\shorttitle{Deuterium fractionation in Perseus}
\author{R. K. Friesen\altaffilmark{1,2}}
\altaffiltext{1}{National Radio Astronomy Observatory, 520 Edgemont Rd, Charlottesville VA 22903}
\altaffiltext{2}{now at the Dunlap Institute for Astronomy and Astrophysics, University of Toronto, 50 St. George St., Toronto, Ontario, Canada, M5S 3H4}
\email{friesen@di.utoronto.ca}

\author{H. M. Kirk \altaffilmark{2}}
\altaffiltext{2}{Origins Institute, McMaster University, 1280 Main Street West, Hamilton, Ontario, Canada, L8S 4M1}

\author{Y. L. Shirley \altaffilmark{3}}
\altaffiltext{3}{Steward Observatory, University of Arizona, 933 N. Cherry Ave.,Tucson, AZ 85721}

\begin{abstract}

We have performed a pointed survey of \ddia\, 2-1 and \ddia\, 3-2 emission toward 64 \dia-bright starless and protostellar cores in the Perseus molecular cloud using the Arizona Radio Observatory Submillimeter Telescope and Kitt Peak 12\,m telescope. We find a mean deuterium fractionation in \dia, $R_D = N(\mbox{\ddia})\,/\,N(\mbox{\dia})$, of 0.08, with a maximum $R_D = 0.2$. In detected sources, we find no significant difference in the deuterium fractionation between starless and protostellar cores, nor between cores in clustered or isolated environments. We compare the deuterium fraction in \dia\, with parameters linked to advanced core evolution. We only find significant correlations between the deuterium fraction and increased H$_2$ column density, as well as with increased central core density, for all cores. Towards protostellar sources, we additionally find a significant anti-correlation between $R_D$ and bolometric temperature. We show that the Perseus cores are characterized by low CO depletion values relative to previous studies of star forming cores, similar to recent results in the Ophiuchus molecular cloud. We suggest that the low average CO depletion is the dominant mechanism that constrains the average deuterium fractionation in the Perseus cores to small values. While current equilibrium and dynamic chemical models are able to reproduce the range of deuterium fractionation values we find in Perseus, reproducing the scatter across the cores requires variation in parameters such as the ionization fraction or the ortho- to para-H$_2$ ratio across the cloud, or a range in core evolution timescales. 

\end{abstract}


\section{Introduction}
\label{sec:intro}

Prior to the formation of a star, the cold ($T \lesssim 20$\,K) and dense ($n \gtrsim 10^5$\,\cc) conditions within star-forming molecular cloud cores drive a cold-gas chemistry that has been well-studied in recent years. Many molecular species, including CO and its isotopologues, become depleted in the gas phase in core centers by freezing out onto dust grains, forming an icy mantle. Depletion in CO has been documented in a number of sources \citep[see, e.g.][]{kuiper96,willacy98,caselli99,bergin01,juvela02,bergin02}, while \citet{christie12} have performed the first systematic survey of CO depletion toward cores embedded in four nearby molecular clouds: Taurus, Serpens, Ophiuchus, and Orion. Being significantly absent from the gas, depleted species are thus no longer good tracers of core physical conditions, and the kinematics revealed by line profiles relate to the core outer envelope only. Molecular species which deplete at higher densities, such as the nitrogen-bearing species \amm\, and \dia, are better suited to probe the structure and kinematics of the interiors of dense cores. Many deuterated molecules, where one or more hydrogen atoms in a molecule is replaced by deuterium, are also preferentially formed in cold core interiors, making them excellent tracers of physical conditions. 

In cold cores, abundances of deuterated molecular species can become enhanced relative to their undeuterated counterparts through reactions with deuterated H$_3^+$, \hd. \hd\, becomes more abundant at cold temperatures through the reaction \citep{millar89}

\begin{equation}
\mbox{H}_3^+ + \mbox{HD} \leftrightarrow \mbox{H}_2\mbox{D}^+ + \mbox{H}_2 + \Delta E
\label{eqn:h2d}
\end{equation}

\showrev{where $\Delta E = 232$\,K when all reactants are in their ground state.}  Equation \ref{eqn:h2d} is thus exothermic in the forward direction, but $\Delta E$ is sufficiently small that the reaction proceeds in both directions at temperatures $T \gtrsim 20$\,K, resulting in no net increase in \hd. In cold cores, however, the energy barrier is sufficient to significantly lessen the backward reaction rate. \hd\, thus can quickly become abundant relative to H$_3^+$, to levels orders of magnitude above the interstellar $[\mbox{D}] / [\mbox{H}]$ ratio of $\sim 10^{-5}$ \citep{york}.  The depletion of CO from the gas phase in cold cores removes an important \hd\, destruction pathway, allowing further enhancement of the deuterium fraction in H$_3^+$.

\showrev{H$_2$ and \hd\, exist in two states, ortho- and para-, depending on the spin of its two hydrogen atoms. The amount of \hd\, in the gas depends strongly on the ortho-to-para-H$_2$ ratio. For example, the backwards reaction of Equation \ref{eqn:h2d} proceeds more rapidly when the main reactants are ortho-\hd\, and ortho-H$_2$. While the equivalent reaction with para-\hd\, and para-H$_2$ is endothermic and slow in cold regions, para-\hd\, is efficiently converted to ortho-\hd\, through reactions with ortho-H$_2$. The overall deuteration of H$_3^+$ (and consequently of \dia) will thus be reduced where the ortho-to-para-H$_2$ ratio is high \citep{pagani92,gerlich02,flower06}. The deuterium chemistry is further complicated by additional parameters which may vary between star forming regions, including the grain size distribution \citep[where depletion onto grains decreases with increasing grain size;][]{caselli08}, and the ionization fraction in the dense gas (which leads to a greater dissociative recombination rate of \hd). }

When \hd\, is abundant, additional reactions further propagate the deuteration abundance enhancement to molecules and molecular ions. Because of the need for high densities, cold temperatures, and depletion of CO, the enhanced deuterium fractionation of molecules typically present in dense cloud cores is thus thought to indicate a dense core is more evolved, and perhaps more likely to be significantly centrally condensed and gravitationally unstable. Indeed, chemical models predict the deuterium fractionation in \dia, for example, is expected to increase with time as the central density of the core increases \citep{aikawa05}. 

Observations of \dia\, and \ddia\, toward star forming regions typically measure the deuterium fractionation, $R_D$, as the ratio of column densities, $R_D = N(\mbox{\ddia})/N(\mbox{\dia})$. Results in low-mass starless and protostellar cores range between $R_D \lesssim 0.1$ \citep{tine00,turner01,crapsi04,belloche06,friesen10b} and $R_D \lesssim 0.5$ \citep{tine00,caselli02,crapsi05,daniel07,bourke12}. Very high $R_D \sim 0.7 - 1$ values have also been determined for several objects, both in beam-averaged observations \citep{miettinen12}, or through radiative transfer modeling \citep{pagani07}. 

Several surveys have looked for trends in the deuterium fractionation in \dia\, with core parameters linked to evolutionary state, such as line width and line asymmetry, central density, temperature and level of CO depletion. In a study of twelve low-mass protostellar and five starless cores, \citet{roberts07} found no systematic difference in $R_D$ between the two samples, and no significant trends in core physical parameters beyond a tendency for the starless cores to have narrower line widths. \citet{crapsi05} surveyed 31 low mass starless cores, and found that while strong correlations were not found between $R_D$ and many parameters, there was a set of properties that, taken together, were associated with the most evolved cores. These `evolved core' properties include the column densities of \dia\, and \ddia, the deuterium fractionation of \dia, and the amount of CO depletion. The \citeauthor{crapsi05} sample included cores from several different nearby star forming regions, including the Perseus and Taurus molecular clouds. Analysed separately, they found the subsample of Taurus cores presented a more homogenous set. In a study of twenty low mass protostellar cores, selected from multiple molecular clouds, \citet{emprechtinger09} found correlations between the deuterium fractionation with dust temperature and CO depletion, where colder, more CO-depleted cores had greater $R_D$ values. The authors also showed that cores with high $R_D$ values were likely to show infall asymmetries in their spectra. These trends were strongest, however, when the sample was limited to the thirteen sources in the Perseus molecular cloud. In a sample of dark clouds, \citet{daniel07} show that the deuterium fractionation in \dia\, is efficient at densities $n \gtrsim 2-3 \times 10^5$\,\cc. In contrast to the trends found by \citeauthor{emprechtinger09} for the Perseus cores, \citeauthor{daniel07}  found no trend in the column density ratio of \ddia\, to \dia\, with temperature in their homologous sample. Comparing results from studies focusing on cores in single regions versus those in different environments suggests that the external environment plays a role in the chemical evolution of star-forming cores. 

Here, we present the first systematic pointed study of the deuterium fractionation in \dia\, toward a large sample of starless and protostellar cores within a single molecular cloud, Perseus, to characterize those factors which impact the core chemistry while excluding cloud-to-cloud variation. We have observed \ddia\, 2-1 and 3-2 toward 64 dense cores in the nearby \citep[250\,pc; e.g.,][]{cernis93} Perseus molecular cloud, selected based on their \dia\, 1-0 brightness in the catalogue of \citet{kirk07}. In addition, many of the targets have been previously surveyed in infrared \citep{evans09}, submillimeter \citep{kirk06} and millimeter \citep{enoch06} continuum emission, as well as in multiple molecular lines, including CO \citep{kirk07,ridge06} and \amm\, \citep{rosolowsky08}. 

We present the source selection and observations in \S \ref{sec:obs}, and describe the spectra and analysis in \S \ref{sec:anal}. While we find several significant correlations between the \dia\, deuterium fraction with parameters expected to be indicators of evolved star-forming cores, including with H$_2$ column density and core central density, there is significant scatter in the trends. Other parameters, such as the kinetic gas temperature, $T_K$, show no correlation with $R_D$, in contrast with the results of several previous studies. Some of the observed scatter may be due to variations in the filling factor of the \ddia\, emission within our beam, particularly for protostellar sources, but we show that this effect should be relatively small. On average, we find that the Perseus cores have low CO depletion, but the few cores with high CO depletion also show higher \dia\, deuterium fractions, as expected from chemical models. For the protostellar cores, we see a significant anti correlation between the bolometric temperature of the protostar and the \dia\, deuterium fraction. In addition, the protostellar sources least likely to be detected in \ddia\, emission also have low envelope masses, showing that the deuterium fraction in protostellar sources decreases as protostars age, heating their surrounding envelopes, and losing envelope mass through radiation, stellar winds, and molecular outflows. In \S \ref{sec:disc} we discuss the deuterium fractionation results in the Perseus cores with respect to the physical properties of the cores and embedded protostars, and compare our findings with chemical models, finding that current equilibrium and dynamic chemical models are able to reproduce the range of deuterium fractionation values we find in Perseus, but require variation in parameters such as the ionization fraction or the ortho- to para-H$_2$ ratio across the cloud, or a range in core evolution timescales, to reproduce the scatter across the Perseus cores. We summarize our results in \S\ref{sec:summary}. 

\begin{deluxetable*}{crrrrrrccc}
\tablecolumns{10}
\tablecaption{Target 850\,\micron\, and \dia\, properties}
\tablehead{
\colhead{ID\tablenotemark{a}} & \colhead{R.A.} & \colhead{decl.} & \colhead{v$_{lsr}$\tablenotemark{b}} & \colhead{$\Delta v$\tablenotemark{b}} &
\colhead{$\int T\Delta v$\tablenotemark{b}} & \colhead{$T_{ex}$} & 
\colhead{SCUBA\tablenotemark{c}} & \colhead{$S_{850}$\tablenotemark{c}} & \colhead{Protostellar?\tablenotemark{d}} \\
\colhead{} & \colhead{J2000} & \colhead{J2000} & \colhead{\kms} & \colhead{\kms} & 
\colhead{K\,\kms} & \colhead{K} & \colhead{} & \colhead{Jy} & \colhead{} }
\startdata
  3  &  3 47  42.3  & 32 52  28.1  &  10.22  &   0.25  &   2.4  &  4.6 (0.5) &   \ldots  &   0.12  &  \ldots \\
  4  &  3 47  41.8  & 32 51  40.3  &  10.24  &   0.45  &   7.4  &  6.2 (0.3) &        1  &   0.59  &       Y \\
  5  &  3 47  40.4  & 32 54  39.0  &  10.20  &   0.26  &   1.9  &  4.7 (1.4) &   \ldots  &   0.04  &  \ldots \\
  6  &  3 47  39.0  & 32 52  11.2  &  10.37  &   0.41  &   4.4  &  5.0 (0.5) &        3  &   0.47  &  \ldots \\
 13  &  3 47  05.6  & 32 43  12.0  &  10.99  &   0.29  &   2.4  &  5.2 (1.2) &   \ldots  &   0.00  &  \ldots \\
 15  &  3 44  48.9  & 32  00  31.8  &   8.94  &   0.29  &   3.0  &  5.1 (0.7) &        4  &   0.37  &  \ldots \\
 19  &  3 44  06.6  & 32  02   05.6  &   8.55  &   0.35  &   4.8  &  6.7 (0.6) &        8  &   0.48  &  \ldots \\
 21  &  3 44   01.8  & 32  01  54.6  &   9.02  &   0.65  &   6.4  &  7.9 (1.1) &       10  &   0.71  &       Y \\
 23  &  3 43  58.2  & 32  04   01.4  &   8.27  &   0.41  &   4.0  & 5.6 (1.4) &       11  &   0.58  &  \ldots \\
 25  &  3 43  57.0  & 32  00  49.7  &   9.06  &   0.48  &   6.9  &  6.3 (0.5) &       13  &   1.99  &       Y \\
 26  &  3 43  51.1  & 32  03  20.9  &   8.56  &   0.42  &   7.3  &  7.6 (0.5) &       15  &   0.87  &       Y \\
 27  &  3 43  44.0  & 32  02  46.4  &   8.24\tablenotemark{e}  &   0.34  &   3.2  &  6.6 (0.1) &   \ldots  &   0.44  &  \ldots \\
 36  &  3 41  43.8  & 31 57  22.0  &   9.39\tablenotemark{e}  &   0.20  &   2.6  &  6.1 (1.1) &   \ldots  &   0.08  &  \ldots \\
 71  &  3 33  21.7  & 31  07  22.2  &   6.66  &   0.87  &  14.1  &  6.1 (0.2) &       29  &   2.44  &       Y \\
 72  &  3 33  17.9  & 31  09  27.8  &   6.27  &   0.76  &  13.4  &  7.0 (0.2) &       30  &   2.60  &       Y \\
 73  &  3 33  16.1  & 31  06  51.6  &   6.43  &   0.56  &  10.7  &  5.5 (0.0) &       32  &   1.41  &       Y \\
 74  &  3 33  13.8  & 31 19  51.3  &   6.84  &   0.32  &   4.5  &  5.0 (0.3) &       33  &   0.46  &       Y \\
 75  &  3 33   05.9  & 31 04  56.6  &   6.63  &   0.44  &   6.2  &  4.8 (0.3) &       34  &   0.46  &  \ldots \\
 77  &  3 32  58.0  & 31  03  44.0  &   6.64  &   0.32  &   4.3  &  5.1 (0.5) &   \ldots  &   0.36  &  \ldots \\
 79  &  3 32  43.2  & 30 59  60.0  &   6.82  &   0.38  &   4.0  &  4.4 (0.5) &   \ldots  &   0.35  &  \ldots \\
 84  &  3 32  28.6  & 31  02  35.0  &   6.61  &   0.30  &   1.6  & 20.9 (3.8) &   \ldots  &   0.22  &  \ldots \\
 85  &  3 32  27.4  & 30 59  22.0  &   6.41  &   0.41  &   3.4  &  3.8 (0.1) &   \ldots  &   0.30  &  \ldots \\
 86  &  3 32  18.0  & 30 49  45.4  &   6.90  &   0.60  &   9.6  &  6.7 (0.3) &       37  &   2.28  &       Y \\
 90  &  3 31  21.0  & 30 45  25.6  &   6.93  &   0.38  &   7.8  &  7.2 (0.4) &       38  &   1.04  &       Y \\
 92  &  3 30  32.0  & 30 26  24.0  &   6.08  &   0.37  &   6.1  &  4.9 (0.1) &   \ldots  &   0.22  &  \ldots \\
 94  &  3 30  15.0  & 30 23  45.0  &   5.90  &   0.38  &   4.7  &  7.8 (1.6) &       39  &   0.26  &  \ldots \\
 95  &  3 29  52.0  & 31 39   03.4  &   8.15  &   0.35  &   6.3  &  6.0 (0.4) &       40  &   0.51  &       Y \\
 96  &  3 29  25.1  & 31 28  16.1  &   7.54  &   0.49  &   5.6  &  5.1 (0.4) &       41  &   0.37  &  \ldots \\
 97  &  3 29  23.4  & 31 33  15.7  &   7.50  &   0.30  &   5.9  &  6.5 (0.4) &       42  &   0.48  &       Y \\
 99  &  3 29  18.4  & 31 25   02.7  &   7.54  &   0.17  &   3.4  &  5.2 (0.3) &       45  &   0.61  &  \ldots \\
100  &  3 29  17.2  & 31 27  44.4  &   7.52  &   0.32  &   5.3  &  7.2 (0.6) &       46  &   0.48  &       Y \\
102  &  3 29  13.2  & 31 13  55.8  &   7.84  &   0.60  &  10.1  &  6.9 (0.3) &       48  &   1.23  &       Y \\
103  &  3 29  11.3  & 31 13   07.4  &   7.15\tablenotemark{e}  &   1.25  &   7.9  &  7.8 (4.1) &       49  &   5.86  &       Y \\
104  &  3 29  10.5  & 31 18  25.1  &   8.59  &   0.53  &  10.4  &  8.0 (0.4) &       50  &   2.13  &       Y \\
106  &  3 29   09.9  & 31 13  31.1  &   7.49\tablenotemark{e}  &   0.97  &  11.4  &  7.1 (1.3) &       52  &  10.28  &       Y \\
107  &  3 29   08.8  & 31 15  12.8  &   7.93  &   0.74  &  16.1  &  7.0 (0.1) &       53  &   1.44  &  \ldots \\
109  &  3 29   06.8  & 31 17  18.3  &   8.48  &   0.53  &   6.5  &  7.3 (0.8) &       56  &   0.64  &  \ldots \\
110  &  3 29   06.5  & 31 15  36.3  &   7.97  &   0.91  &  16.6  & 11.5 (0.6) &       57  &   1.63  &  \ldots \\
111  &  3 29   03.7  & 31 14  47.7  &   6.82\tablenotemark{e}  &   0.80  &  17.7  &  10.7 (2.2) &       60  &   1.05  &       Y \\
112  &  3 29   03.2  & 31 15  53.6  &   8.40  &   0.68  &  18.5  & 10.1 (0.3) &       61  &   5.08  &       Y \\
113  &  3 29   01.4  & 31 20  23.1  &   7.97  &   0.60  &  11.8  &  7.9 (0.3) &       62  &   2.14  &       Y \\
115  &  3 29   00.2  & 31 11  53.0  &   7.16  &   0.64  &   8.5  &  7.3 (0.5) &       64  &   0.26  &       Y \\
116  &  3 28  59.5  & 31 21  28.7  &   7.83  &   0.77  &   9.9  &  9.1 (1.0) &       65  &   1.47  &       Y \\
118  &  3 28  55.3  & 31 14  27.9  &   7.34\tablenotemark{e}  &   0.58  &  12.4  &  6.3 (0.5) &       66  &   2.92  &       Y \\
121  &  3 28  42.5  & 31  06  13.1  &   7.22  &   0.38  &   6.0  &  6.1 (0.4) &       67  &   0.39  &  \ldots \\
122  &  3 28  40.1  & 31 17  48.4  &   7.98  &   0.53  &  10.3  &  6.5 (0.2) &       68  &   0.84  &       Y \\
123  &  3 28  39.1  & 31 18  24.1  &   8.19  &   0.48  &   9.1  &  5.7 (0.2) &       69  &   0.82  &  \ldots \\
124  &  3 28  38.8  & 31 05  54.3  &   7.04  &   0.34  &   6.3  &  6.2 (0.4) &       70  &   0.42  &       Y \\
125  &  3 28  36.7  & 31 13  23.6  &   7.28  &   0.50  &   5.8  & 6.4 (1.6) &       71  &   0.88  &       Y \\
126  &  3 28  34.5  & 31  06  59.3  &   6.84  &   0.31  &   3.5  &  6.2 (1.1) &       72  &   0.23  &       Y \\
128  &  3 28  32.3  & 31 10  58.7  &   7.21  &   0.38  &   4.8  &  5.3 (0.4) &       73  &   0.32  &       Y \\
130  &  3 27  57.1  & 30  07  57.0  &   4.85  &   0.39  &   1.4  &  4.0 (2.6) &   \ldots  &   0.05  &  \ldots \\
131  &  3 27  55.6  & 30  05  40.0  &   4.78  &   0.25  &   1.2  & 19.5 (4.1) &   \ldots  &   0.00  &  \ldots \\
133  &  3 27  48.3  & 30 12   08.0  &   5.06  &   0.55  &   6.8  &  6.0 (0.5) &       75  &   0.60  &       Y \\
134  &  3 27  42.7  & 30 12  24.5  &   4.87  &   0.48  &   9.5  &  7.0 (0.1) &       76  &   0.78  &       Y \\
135  &  3 27  39.0  & 30 12  53.6  &   4.72  &   0.62  &   9.4  &  6.0 (0.3) &       78  &   0.89  &       Y \\
136  &  3 27  37.9  & 30 13  53.2  &   5.94\tablenotemark{e}  &   0.51  &   6.6  &  7.9 (1.1) &       79  &   0.40  &       Y \\
143  &  3 26  37.2  & 30 15  18.7  &   5.09  &   0.29  &   4.3  &  5.1 (0.4) &       80  &   0.37  &       Y \\
146  &  3 25  49.3  & 30 42  15.1  &   4.49  &   0.37  &   6.2  &  5.3 (0.3) &       82  &   0.65  &  \ldots \\
147  &  3 25  46.3  & 30 44  14.1  &   4.64  &   0.47  &   4.3  &  5.1 (0.6) &   \ldots  &   0.16  &  \ldots \\
148  &  3 25  38.9  & 30 43  59.8  &   5.03\tablenotemark{e}  &   0.90  &  13.0  &  6.6 (0.5) &       83  &   2.00  &       Y \\
149  &  3 25  36.0  & 30 45  10.8  &   4.51  &   0.82  &  20.2  &  8.4 (0.2) &       84  &   4.36  &       Y \\
150  &  3 25  25.7  & 30 45   01.6  &   4.05  &   0.49  &   9.4  &  5.4 (0.0) &       85  &   0.83  &  \ldots \\
152  &  3 25  22.5  & 30 45   06.5  &   4.08  &   0.50  &  10.9  &  5.8 (0.2) &       86  &   1.64  &       Y \\
\enddata
\label{tab:sourceInfo}
\tablenotetext{a}{\dia\, core ID numbers.}
\tablenotetext{b}{\dia\, $v_{lsr}$, FWHM $\Delta v$, $\tau \times (T_{ex} - T_{bg})$, and $T_{ex}$ based on hyperfine spectral line fitting. The tdv function was used to calculate  $\tau \times (T_{ex} - T_{bg})$.}
\tablenotetext{c}{SCUBA ID number and total SCUBA flux within a 30\arcsec\, FWHM beam.}
\tablenotetext{d}{Protostellar designation as discussed in the text.}
\tablecomments{Units of right ascension are hours, minutes, and seconds, and units of declination are degrees, arcminutes, and arcseconds. All core ID numbers and \dia\, emission line properties are taken from \citet{kirk07}. The SCUBA data were originally described in \citet{kirk06}, and have been convolved to a 30\arcsec\, FWHM for this analysis. The designation of a core as protostellar relies on multiple literature sources, described in detail in \S\ref{sec:targets}.}
\tablenotetext{e}{Sources where two velocity components along the line of sight are identified in \dia\, 1-0 line profiles. }
\end{deluxetable*}

\section{Source data}
\label{sec:obs}

\subsection{Source selection}
\label{sec:targets}

The observed \ddia\, cores were selected based on their bright \dia\, 1-0 emission from \citet{kirk07}, but nevertheless span a wide range of conditions in the Perseus molecular cloud. The \citeauthor{kirk07} sample included candidate dense cores from submillimeter data, points of high visual extinction from Palomar plates, and peaks of large-scale extinction from 2MASS data. Within our subset of 64 sources, fifty-one are located at the peak of an 850\,\micron\, continuum clump. Some core properties (Right Ascension, Declination, velocity of the local standard of rest, $v_{LSR}$, FWHM line width, $\Delta v$, and integrated intensity of associated \dia\, 1-0 emission, as well as SCUBA flux and identification as protostellar or starless) are listed in Table \ref{tab:sourceInfo}. Based on submillimeter continuum measurements, the cores in our sample have masses ranging from $\sim 0.1 $M$_\odot$ to $\sim 10 $M$_\odot$ (median $M = 1.4\,M_\odot$), effective angular radii from $\sim 15$\arcsec\, to $\sim 55$\arcsec\, (median $r = 28$\arcsec; 0.03\,pc or 7000\,AU at 250\,pc), and have a median average density of $\sim 2 \times 10^5$\,\cc\, \citep{kirk06}. 

To identify cores as starless or protostellar, we begin with the classification of \citet{kirk07}, which is based on analysis of Spitzer data in Perseus by \citet{jorgensen07}. We further compare our target list with the catalogue of \citet{enoch09}, who identified protostellar sources using Spitzer and 1\,mm Bolocam data, and classify as protostellar those cores which have a protostellar source within 20\arcsec\, of the pointing location. Three additional Perseus cores in our sample show evidence of harboring embedded protostellar objects in recent studies based on molecular line or deep continuum observations \citep{schnee12,chen10,enoch10,schnee10}, despite having been previously classified as `starless' in the studies described above. Including these sources in our protostellar subset, we find that a greater fraction of the Perseus cores are classified as protostellar (43/64) than starless (21/64). We further classify cores as `clustered' if they reside within the NGC 1333 or IC348 regions, following the boundary definitions of each region given by \citet{jorgensen06}. The targets are evenly divided between those found in the clustered star-forming regions (32/64), and those in more isolated parts of the cloud (32/64). The more isolated cores nonetheless reside largely in well-known groupings within Perseus, including the B5, B1, and L1448 regions. 

\begin{figure*}[t]
\plotone{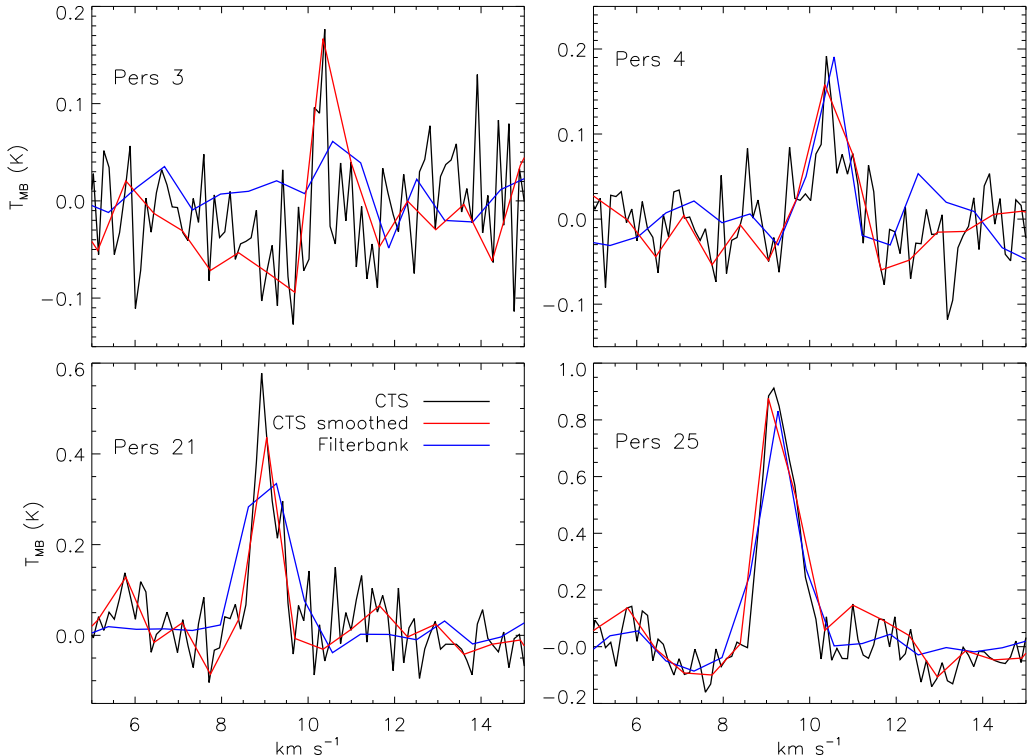}
\caption{Comparison of \ddia\, 3-2 spectra obtained toward four Perseus sources with the CTS (black lines; light grey lines show the CTS data smoothed to the filter bank resolution) and the filterbank backends (grey lines; red in the online version). \label{fig:spectra}}
\end{figure*}

\subsection{SMT observations of \ddia\, 3-2}

Sixty-four Perseus cores were observed at the Arizona Radio Observatory Submillimeter Telescope (SMT) in 2010 April 10-11, November 27-28, and 2011 February 16. During the April epoch, the observations were performed using the ALMA Band 6 prototype receiver (1\,mm) with the Chirp Transform Spectrometer (CTS) as the backend. The spectral resolution was 47 kHz, corresponding to 0.06\,\kms. Most sources were observed for three 4 minute scans in position-switching mode, with additional integration time on several weak sources to increase the signal to noise ratio. Pointing observations were performed every 1-1.5 hours by offsetting the detector frequency to the $^{12}$CO 2-1 transition (230.5\,GHz).  During the later two sets of observations, the CTS was not available, and the filterbank backend was used instead. The spectral resolution was 250 kHz, corresponding to 0.32\.\kms. All sources observed with the spectrometer were re-observed with the filterbanks. Each source was typically observed for two 5 minute scans on both the November and February runs, with several final additional scans on several low signal-to-noise (S/N) detections in the November run. The SMT beam at 231\,GHz is $\sim 30$\arcsec\, (FWHM), well-matched to both the \dia\, (26\arcsec\, FWHM) and \amm\, (32\arcsec\, FWHM) data.

All spectra were reduced in CLASS\footnotemark\footnotetext{CLASS is part of the GILDAS software suite available from 
http://www.iram.fr/IRAMFR/GILDAS}.  We first flattened each spectrum using a 4th order polynomial.  Several scans had highly irregular baselines which could not be fit with the polynomial; these were discarded from further use, and, where possible, re-observed.

The spectra were calibrated for each observing run based on observations of Mars (data taken in April) and Jupiter (data taken in November and February). In November, the vertical (V) polarization data showed inconsistent results in calibration scans, and these data were omitted from the final dataset. The calibration efficiencies were found to be $0.82 \pm 0.07$ for the April run, $0.68 \pm 0.02$ for the horizontal (H) polarization of the November run, and $0.55 \pm 0.02$ and $0.48 \pm 0.02$ for the H and V polarizations of the February run. \showrev{The change in calibration efficiency between November and February may have been caused by a warmup of the receiver between the two epochs. Several Perseus targets with strong \ddia\, detections were observed in more than one run, to allow for a secondary check of the beam efficiencies derived. The relative efficiencies were found to be in agreement with the planet-derived values, within uncertainties of $\sim 15$\,\% in the observed peak line intensities. }

After applying the calibration factors, all spectra of each source were summed together. We show in Figure \ref{fig:spectra} a comparison of the \ddia\, 3-2 spectra toward four sources which were observed with both the CTS and filterbank backends. The Figure shows that the filterbank data recover the line emission for both faint and strong emission lines, but sources with narrow lines will suffer from artificial line broadening and depressed peak line brightnesses due to the poor spectral resolution of the filterbanks. In particular, cores with both faint and narrow \ddia\, emission that would be detected by the CTS observations may not be detected in the filterbank data (see Pers 3 in the Figure). Of the eleven sources observed with both the CTS and the filterbanks, only one source (Pers 3) showed line emission in the CTS data but not in the filterbank data. For this source, we include only the CTS observations in the final data. Our analysis below depends only on the \ddia\, 3-2 line centroids and integrated intensity, which agree well between the CTS and filterbank data, and we thus conclude that our results are not significantly affected by the low spectral resolution of the data. 

The \ddia\, 3-2 emission line contains hyperfine structure, but no structure was visible at a 3$\sigma$ level above the rms noise in the data for any Perseus sources (Pers 25 shows a marginal detection of the satellite lines). A single Gaussian fit in CLASS was thus used to characterize the emission, and the returned area from the fit is reported as the line integrated intensity, $W_{32}$, for sources detected with S/N $> 3$. To test the effects on our results from the low velocity resolution of the filterbank data, and the fitting of a single gaussian rather than the full hyperfine structure to the lines, we additionally performed hyperfine fits toward those sources observed at high velocity resolution with the CTS. We find that the line centroids are recovered by the single gaussian within the filterbank velocity resolution of 0.32\,\kms, but the returned gaussian line widths are a factor $\sim 2$ larger than those found by the hyperfine fits. \showrev{The returned area from the gaussian and hyperfine fits additionally agree within uncertainties. } Final sensitivities for each source are listed in Table \ref{tab:sourceDetections}. 

\subsection{Kitt Peak 12\,m Telescope observations of \ddia\, 2-1}

Observations of \ddia\, 2-1 were made toward 54 Perseus cores within the allotted time at the Arizona Radio Observatory Kitt Peak 12\,m over 2010 November 29 and 30, and December 11, 12 and 13. Using the 2\,mm receiver, the observation rest frequency was set to 151.217178\,GHz, the frequency of the strongest hyperfine component of the 2-1 transition. The backend used was the Millimeter Autocorrelator, with 100\,MHz total bandwidth and 6.1\,kHz spectral resolution (0.012\,\kms\, at 151.2\,GHz). Sources were observed in position-switching mode for two or three periods of six minutes each to achieve similar rms noise levels based on weather conditions. The data were corrected at the telescope for the atmosphere and for telescope losses. \showrev{The aperture efficiency of the telescope is 45\% at 145 GHz\footnotemark{}\footnotetext{http://aro.as.arizona.edu/12\_obs\_manual/chapter\_3.htm}. }Further calibration to the $T_{MB}$ scale was done using observations of Jupiter, which were performed once each shift. Based on the Jupiter measurements, uncertainty in the final calibration scale is $\sim 5$\%. At 151\,GHz, the telescope beam is $\sim 41$\arcsec\, (FWHM). 

Spectra were reduced in CLASS. The data were smoothed to a final velocity resolution of 0.14\,\kms. A linear baseline was removed and the data were summed for each source. Detections were fit using the hyperfine structure (hfs) CLASS routine, using line frequencies and relative strengths calculated from \citet{pagani09}. The resulting line centroid velocities, $v_{LSR}$, and widths, $\Delta v$, should have better accuracy than would be provided by a single Gaussian fit that did not take into account the hyperfine structure of the line. We do not discuss further the derived \ddia\, 2-1 centroid velocities, however, due to an unexplained, episodic velocity shift observed in some of the data. While this observed velocity shift in the observed line, when present, was of a similar magnitude in nearly all clearly detected sources, there was no clear pattern in time when the offset occurred, and no obvious cause in the telescope log. We find no evidence in \ddia-bright sources that the velocity shift occurred between scans of the same source, so are confident the resulting summed data accurately represent the source line emission. In nearly all sources, the returned line opacities also have large uncertainties due to a lack of high S/N in the satellite components. We therefore primarily use the hfs-fit velocity information to determine three windows over which we measured the integrated \ddia\, 2-1 intensity for each core, $W_{21}$, based on the line $v_{LSR}$ and $\Delta v$, and the expected locations of \ddia\, 2-1 hyperfine components. Where no line was detected, an upper limit to the integrated intensity was determined, following the same method but defining spectral windows based on the centroid velocity and width of the observed \dia\, 1-0 line toward the core. Final sensitivities for each source are listed in Table \ref{tab:sourceDetections}. 

\subsection{Other data}
\label{sec:otherData}
 
In our analysis, we make use of several published catalogs focused on the Perseus molecular cloud. The \dia\, 1-0 catalog used to select cores for \ddia\, observations was based in part on previous 850\,\micron\, continuum observations performed with the James Clerk Maxwell Telescope using the Submillimetre Common User Bolometer Array (SCUBA) at 15\arcsec\, FWHM \citep{kirk06}, smoothed to a final effective resolution of 20\arcsec\, FWHM to reduce pixel-to-pixel noise. In their \dia\, study, \citet{kirk07} also present observations of C$^{18}$O 2-1 made with the IRAM 30\,m telescope at 11\arcsec\, FWHM. All cores with \dia\, emission were also detected in C$^{18}$O. The C$^{18}$O emission was comprised of multiple velocity components for many cores, and in their analysis, \citeauthor{kirk06} determine the individual C$^{18}$O component that best matches the \dia\, emission for each core. We use these matched components in our analysis and discussion below. 

\citet{rosolowsky08} performed a targeted study of \amm\, (1,1) and (2,2) emission toward a large number of sources in Perseus, calculating kinetic temperatures for each core. Their targets were selected from several catalogs, foremost a 1.1\,mm study of the region, observed using BOLOCAM at the Caltech Submillimeter Observatory \citep{enoch06}. Consequently, the \amm\, observation centers are frequently offset relative to our \dia\, and \ddia\, observations. For our targets, however, most of the offsets remain small ($\lesssim 15$\arcsec, or half the \ddia\, 3-2 beam FWHM at the SMT). Of those Perseus cores detected in \ddia\, emission with offsets of $\sim 20$\arcsec - 25\arcsec\, between the \amm\, and \dia\, targets (five cores), only one shows a moderate ($\sim 0.5$\,\kms) offset in line-of-sight velocity between the \amm\, and \dia\, observations, with the rest in agreement within observational uncertainties. We thus expect that the \dia\, and \amm\, observations are tracing similar material for most of our targets, and that the kinetic temperature measurements from the \amm\, analysis by \citet{rosolowsky08} are accurate for the cores observed in our study. Four cores in our catalogue (23, 71, 125, and 148), which were all detected in \ddia\, emission, have no nearby counterparts in the \amm\, dataset. 

\figsetstart
\figsetnum{2}
\figsettitle{Spectra of N$_2$H$^+$ and N$_2$D$^+$ }

\figsetend

\begin{figure}
\epsscale{1.2}
\plotone{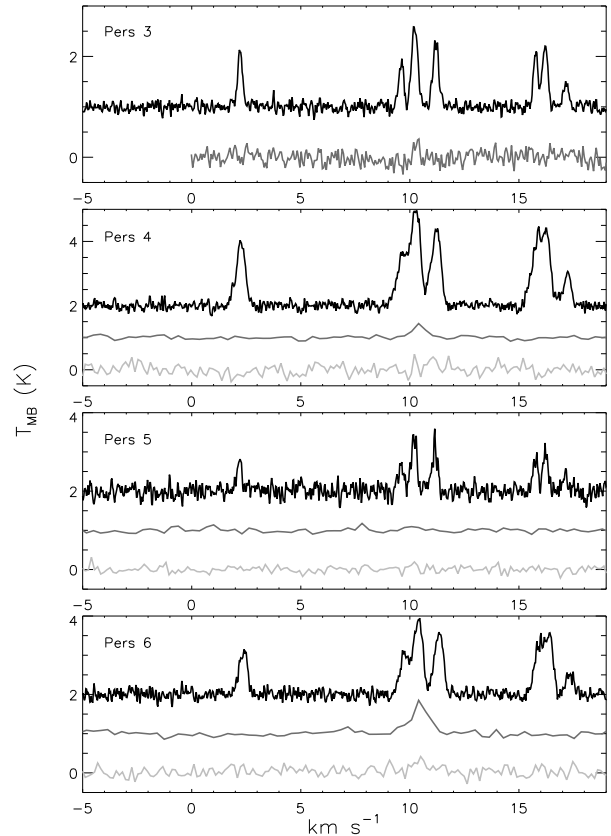}
\caption{Spectra of N$_2$H$^+$ and N$_2$D$^+$  toward each of the 64 Perseus 
cores in this study. N$_2$H$^+$ 1-0 spectra are shown in black, N$_2$D$^+$ 2-1 
spectra are shown in grey (blue in the online version), and N$_2$D$^+$ 3-2 spectra are shown in dark grey (red in the online version). For 
clarity, the N$_2$D$^+$ 3-2 data have been multiplied by a factor of 2, and
spectra are offset from zero to display all available datasets. Some cores were 
observed in one N$_2$D$^+$ transition only. The N$_2$H$^+$ 1-0 data were
first presented in Kirk et al. (2007). \label{fig:allSpec}
}
\epsscale{1}
\end{figure}

\section{Results and analysis}
\label{sec:anal}

We list the parameters of the fits to detected sources in Table \ref{tab:sourceDetections}. For this study, a core is `detected' if emission was seen with S/N $> 3$ in $T_{MB}$ in either the \ddia\, 2-1 or 3-2 transitions. Of the 54 Perseus cores observed in emission from both \ddia\, 2-1 and 3-2, 45 were detected in both transitions with S/N $> 3$. Two sources were detected only in \ddia\, 2-1 emission, four were detected only in \ddia\, 3-2 emission (likely due to insufficient sensitivity in the \ddia\, 2-1 observations), and three were undetected in either transition.  Of the remaining ten cores which were not observed in \ddia\, 2-1 emission, nine were undetected in emission from \ddia\, 3-2. We show in Figure \ref{fig:allSpec} the \dia\, 1-0, \ddia\, 2-1, and \ddia\, 3-2 spectra toward each target. 

We can further break down the detections and non-detections as a function of whether the core has previously been identified as starless or protostellar, and whether the core lies within a clustered region of the Perseus molecular cloud. Based on these definitions, 36 of 43 (84\,\%) protostellar sources were detected, and 16 of 21 (76\,\%) starless sources were detected. Of the clustered sources, 29 of 32 (91\,\%) were detected, and 23 of 32 (72\,\%) isolated sources detected.

\begin{deluxetable*}{rrrrrrr|rrrrr}
\tablecolumns{12}
\tabletypesize{\scriptsize}
\tablecaption{\ddia\, line properties \label{tab:sourceDetections}}
\tablehead{
\colhead{ID} & 
\multicolumn{6}{c}{\ddia\, 2-1} & 
\multicolumn{5}{c}{\ddia\, 3-2} \\ 
\colhead{} & \colhead{$v_{LSR}$} & \colhead{$\Delta v$} &
\colhead{$T\tau$} & \colhead{$\tau$} &
\colhead{rms} & \colhead{$W_{21}$} &
\colhead{$v_{LSR}$} & \colhead{$\Delta v$} &
\colhead{$I_{peak}$} & \colhead{rms} & \colhead{$W_{32}$} \\
\colhead{} & \colhead{\kms} & \colhead{\kms} &
\colhead{K} & \colhead{} & \colhead{K} & \colhead{K\,\kms} &
\colhead{\kms} & \colhead{\kms} & \colhead{K} &
\colhead{K} & \colhead{K\,\kms} }
\startdata
\multicolumn{12}{c}{Detected in both \ddia\, 2-1 and 3-2} \\
\cline {1-12}
  6 & 10.38 (0.09) & 0.60 (0.22) & 1.1 (0.6) & 2.6 (3.1) & 0.13 & 0.77 (0.04) & 10.47 (0.03) & 0.95 (0.09) & 0.38 & 0.03 & 0.38 (0.03) \\
 21 &  8.91 (0.04) & 0.46 (0.09) & 1.6 (0.6) & 2.6 (2.2) & 0.12 & 0.82 (0.04) &  9.00 (0.02) & 0.83 (0.05) & 0.46 & 0.03 & 0.41 (0.02) \\
 23 &  8.36 (0.05) & 0.34 (0.09) & 2.0 (0.8) & 7.4 (4.2) & 0.10 & 0.42 (0.03) &  8.51 (0.14) & 1.38 (0.28) & 0.13 & 0.04 & 0.19 (0.04) \\
 25 &  9.06 (0.03) & 0.87 (0.01) & 2.0 (0.1) & 1.0 (0.0) & 0.11 & 1.83 (0.04) &  9.27 (0.01) & 0.94 (0.03) & 0.93 & 0.03 & 0.94 (0.03) \\
 26 &  8.60 (0.03) & 0.39 (0.07) & 2.6 (0.8) & 3.7 (1.8) & 0.12 & 0.79 (0.04) &  8.70 (0.01) & 0.77 (0.04) & 0.60 & 0.03 & 0.49 (0.02) \\
 27 &  7.78 (0.02) & 0.16 (0.06) & 4.0 (4.0) & 10.2 (2.5) & 0.13 & 0.45 (0.04) &  8.43 (0.05) & 0.69 (0.12) & 0.14 & 0.02 & 0.10 (0.02) \\
 36 & 10.05 (0.06) & 0.34 (0.12) & 0.4 (0.1) & 0.1 (0.1) & 0.09 & 0.14 (0.03) & 13.46 (0.14) & 0.78 (0.35) & 0.12 & 0.05 & 0.10 (0.04) \\
 71 &  7.28 (0.03) & 0.94 (0.08) & 2.7 (0.4) & 3.0 (0.8) & 0.09 & 1.92 (0.03) &  6.84 (0.02) & 1.18 (0.04) & 1.02 & 0.04 & 1.28 (0.04) \\
 72 &  6.29 (0.04) & 0.94 (0.13) & 2.0 (0.5) & 2.3 (1.1) & 0.11 & 1.55 (0.04) &  6.41 (0.01) & 1.05 (0.04) & 0.77 & 0.03 & 0.86 (0.03) \\
 73 &  7.04 (0.04) & 0.63 (0.11) & 1.8 (0.6) & 0.7 (1.4) & 0.14 & 0.98 (0.05) &  6.57 (0.02) & 0.82 (0.03) & 0.88 & 0.04 & 0.77 (0.03) \\
 74 &  6.89 (0.04) & 0.38 (0.07) & 2.3 (0.6) & 5.2 (2.1) & 0.09 & 0.79 (0.03) &  6.91 (0.02) & 0.55 (0.08) & 0.42 & 0.03 & 0.24 (0.03) \\
 75 &  7.05 (0.07) & 0.77 (0.16) & 0.8 (0.1) & 0.1 (1.5) & 0.13 & 0.64 (0.05) &  6.66 (0.09) & 0.97 (0.30) & 0.14 & 0.03 & 0.14 (0.03) \\
 77 &  6.67 (0.04) & 0.24 (0.08) & 3.5 (3.1) & 9.0 (5.3) & 0.12 & 0.44 (0.04) &  6.74 (0.08) & 0.81 (0.22) & 0.18 & 0.04 & 0.16 (0.03) \\
 86 &  7.55 (0.04) & 0.77 (0.12) & 1.9 (0.5) & 1.2 (1.2) & 0.12 & 1.02 (0.04) &  7.11 (0.02) & 0.94 (0.05) & 0.67 & 0.04 & 0.67 (0.03) \\
 90 &  6.98 (0.03) & 0.30 (0.06) & 3.5 (1.0) & 4.0 (2.3) & 0.15 & 0.89 (0.05) &  7.07 (0.02) & 0.80 (0.04) & 0.69 & 0.04 & 0.59 (0.03) \\
 92 &  6.10 (0.04) & 0.31 (0.08) & 2.4 (1.0) & 7.1 (4.2) & 0.12 & 0.62 (0.04) &  6.26 (0.04) & 0.68 (0.12) & 0.27 & 0.03 & 0.20 (0.03) \\
 95 &  8.79 (0.02) & 0.36 (0.05) & 1.6 (0.4) & 0.2 (4.5) & 0.09 & 0.57 (0.03) &  8.20 (0.03) & 0.59 (0.06) & 0.54 & 0.05 & 0.34 (0.03) \\
 96 &  8.13 (0.06) & 0.63 (0.10) & 0.8 (0.1) & 0.1 (1.7) & 0.12 & 0.27 (0.04) &  7.78 (0.03) & 0.72 (0.07) & 0.36 & 0.03 & 0.28 (0.02) \\
 97 &  8.12 (0.02) & 0.36 (0.04) & 2.5 (0.6) & 2.3 (1.2) & 0.09 & 0.85 (0.03) &  7.60 (0.02) & 0.51 (0.06) & 0.54 & 0.03 & 0.30 (0.02) \\
 99 &  7.68 (0.11) & 0.68 (0.25) & 1.0 (0.7) & 1.9 (3.5) & 0.18 & 0.70 (0.06) &  7.09 (0.05) & 0.66 (0.09) & 0.25 & 0.04 & 0.18 (0.02) \\
100 &  8.09 (0.04) & 0.31 (0.09) & 2.1 (1.0) & 3.4 (3.2) & 0.13 & 0.51 (0.04) &  7.63 (0.04) & 0.83 (0.09) & 0.31 & 0.03 & 0.28 (0.03) \\
102 &  7.76 (0.03) & 0.66 (0.07) & 4.0 (0.7) & 3.5 (1.0) & 0.12 & 2.10 (0.04) &  7.91 (0.02) & 0.91 (0.06) & 0.74 & 0.04 & 0.71 (0.04) \\
103 &  7.72 (0.03) & 0.85 (0.13) & 1.9 (0.5) & 1.8 (1.1) & 0.10 & 1.41 (0.04) &  7.04 (0.01) & 0.97 (0.03) & 1.14 & 0.04 & 1.18 (0.03) \\
104 &  9.20 (0.02) & 0.37 (0.04) & 4.4 (0.8) & 6.2 (1.6) & 0.11 & 1.10 (0.04) &  8.78 (0.03) & 0.80 (0.07) & 0.36 & 0.03 & 0.31 (0.02) \\
106 &  7.68 (0.05) & 1.28 (0.20) & 0.9 (0.3) & 0.8 (1.2) & 0.09 & 1.14 (0.03) &  7.14 (0.02) & 1.22 (0.05) & 0.83 & 0.04 & 1.08 (0.04) \\
107 &  8.27 (0.02) & 0.99 (0.05) & 3.8 (0.3) & 2.8 (0.4) & 0.08 & 2.81 (0.03) &  7.91 (0.01) & 1.23 (0.03) & 1.06 & 0.03 & 1.38 (0.03) \\
109 &  8.48 (0.03) & 0.35 (0.06) & 4.1 (1.1) & 6.7 (2.3) & 0.12 & 1.02 (0.04) &  8.64 (0.03) & 0.94 (0.07) & 0.50 & 0.03 & 0.51 (0.03) \\
110 &  8.49 (0.02) & 1.09 (0.07) & 2.3 (0.3) & 1.2 (0.5) & 0.09 & 2.29 (0.03) &  8.10 (0.01) & 1.05 (0.03) & 1.03 & 0.03 & 1.15 (0.03) \\
111 &  6.90 (0.05) & 0.93 (0.22) & 1.2 (0.4) & 1.4 (1.4) & 0.10 & 1.18 (0.04) &  7.05 (0.03) & 1.06 (0.06) & 0.53 & 0.03 & 0.60 (0.03) \\
112 &  8.28 (0.04) & 0.44 (0.09) & 3.2 (0.9) & 3.7 (1.9) & 0.15 & 1.14 (0.05) &  8.49 (0.02) & 0.90 (0.05) & 0.75 & 0.04 & 0.73 (0.03) \\
113 &  7.90 (0.03) & 0.64 (0.09) & 2.4 (0.6) & 1.7 (1.1) & 0.12 & 1.30 (0.04) &  8.11 (0.02) & 0.88 (0.04) & 0.99 & 0.04 & 0.93 (0.04) \\
115 &  7.91 (0.05) & 0.47 (0.14) & 0.9 (0.5) & 1.7 (2.7) & 0.10 & 0.30 (0.03) &  7.19 (0.03) & 0.74 (0.09) & 0.38 & 0.04 & 0.30 (0.03) \\
116 &  8.20 (0.07) & 1.35 (0.16) & 0.6 (0.1) & 0.1 (2.0) & 0.10 & 0.89 (0.04) &  7.91 (0.02) & 0.88 (0.04) & 0.66 & 0.03 & 0.62 (0.03) \\
118 &  7.29 (0.06) & 0.75 (0.20) & 1.0 (0.5) & 0.7 (1.9) & 0.13 & 0.86 (0.05) &  7.55 (0.03) & 0.82 (0.07) & 0.34 & 0.03 & 0.29 (0.02) \\
121 &  7.88 (0.04) & 0.40 (0.09) & 0.9 (0.5) & 0.3 (5.6) & 0.10 & 0.25 (0.03) &  7.26 (0.04) & 0.66 (0.10) & 0.27 & 0.03 & 0.19 (0.02) \\
122 &  8.56 (0.04) & 0.56 (0.10) & 1.2 (0.2) & 0.1 (3.2) & 0.15 & 0.59 (0.05) &  8.15 (0.05) & 0.97 (0.11) & 0.30 & 0.03 & 0.31 (0.03) \\
123 &  8.71 (0.04) & 0.53 (0.15) & 1.4 (0.8) & 0.3 (4.1) & 0.14 & 0.74 (0.05) &  8.15 (0.03) & 0.67 (0.10) & 0.37 & 0.04 & 0.27 (0.03) \\
124 &  7.66 (0.03) & 0.41 (0.08) & 1.4 (0.6) & 1.6 (2.0) & 0.10 & 0.46 (0.03) &  7.15 (0.06) & 0.86 (0.15) & 0.27 & 0.04 & 0.25 (0.03) \\
134 &  4.86 (0.05) & 0.52 (0.09) & 1.1 (0.3) & 0.1 (9.5) & 0.15 & 0.46 (0.05) &  4.96 (0.03) & 0.78 (0.10) & 0.44 & 0.04 & 0.36 (0.03) \\
135 &  3.91 (0.06) & 0.49 (0.10) & 2.1 (0.7) & 8.5 (3.5) & 0.09 & 0.52 (0.03) &  4.73 (0.07) & 1.16 (0.16) & 0.25 & 0.04 & 0.31 (0.04) \\
136 &  6.02 (0.48) & 2.05 (0.74) & 0.3 (0.3) & 6.5 (7.2) & 0.09 & 0.26 (0.04) &  4.66 (0.08) & 0.45 (0.22) & 0.16 & 0.04 & 0.08 (0.03) \\
146 &  3.61 (0.05) & 0.51 (0.10) & 2.3 (0.7) & 5.5 (2.1) & 0.09 & 0.65 (0.03) &  4.65 (0.02) & 0.56 (0.05) & 0.55 & 0.04 & 0.32 (0.03) \\
148 &  4.39 (0.03) & 0.73 (0.10) & 1.6 (0.5) & 0.7 (1.0) & 0.09 & 0.86 (0.03) &  5.37 (0.03) & 0.98 (0.05) & 0.86 & 0.05 & 0.90 (0.05) \\
149 &  3.52 (0.02) & 0.73 (0.05) & 3.8 (0.5) & 2.3 (0.6) & 0.09 & 2.18 (0.03) &  4.60 (0.01) & 1.07 (0.03) & 1.61 & 0.06 & 1.82 (0.05) \\
150 &  3.09 (0.02) & 0.46 (0.05) & 2.5 (0.6) & 3.2 (1.3) & 0.09 & 0.77 (0.03) &  4.17 (0.05) & 0.76 (0.11) & 0.50 & 0.06 & 0.41 (0.05) \\
152 &  3.15 (0.03) & 0.51 (0.06) & 1.1 (0.1) & 0.1 (1.7) & 0.09 & 0.51 (0.03) &  4.34 (0.05) & 0.75 (0.15) & 0.41 & 0.06 & 0.33 (0.05) \\
\cline {1-12}
\multicolumn{12}{c}{Detected in \ddia\, 2-1 only} \\
\cline {1-12}
 36 & 10.05 (0.06) & 0.34 (0.12) & 0.4 (0.1) & 0.1 (0.1) & 0.09 & 0.14 (0.03) & \nodata & \nodata & \nodata & 0.05 & $\le$ 0.04 \\
147 &  3.73 (0.07) & 0.54 (0.15) & 0.7 (0.4) & 1.6 (2.8) & 0.08 & 0.33 (0.03) & \nodata & \nodata & \nodata & 0.05 & $\le$ 0.11 \\
\cline {1-12}
\multicolumn{12}{c}{Detected in \ddia\, 3-2 only} \\
\cline {1-12}
  3 & \nodata & \nodata & \nodata & \nodata & \nodata & \nodata & 10.32 (0.03) & 0.28 (0.05) & 0.19 & 0.06 & 0.06 (0.01) \\
  4 & \nodata & \nodata & \nodata & \nodata & 0.15 & $\le$ 0.34 & 10.42 (0.04) & 0.68 (0.10) & 0.22 & 0.02 & 0.16 (0.02) \\
 15 & \nodata & \nodata & \nodata & \nodata & 0.11 & $\le$ 0.29 &  9.10 (0.08) & 0.77 (0.18) & 0.11 & 0.03 & 0.09 (0.02) \\
 19 & \nodata & \nodata & \nodata & \nodata & 0.08 & $\le$ 0.06 &  8.51 (0.07) & 0.49 (0.17) & 0.14 & 0.03 & 0.07 (0.02) \\
143 & \nodata & \nodata & \nodata & \nodata & 0.08 & $\le$ 0.19 &  5.45 (0.08) & 0.60 (0.20) & 0.17 & 0.04 & 0.11 (0.03) \\
\enddata
\end{deluxetable*}

\subsection{Kinematics}
\label{sec:kin}

\begin{figure*}[t]
\plottwo{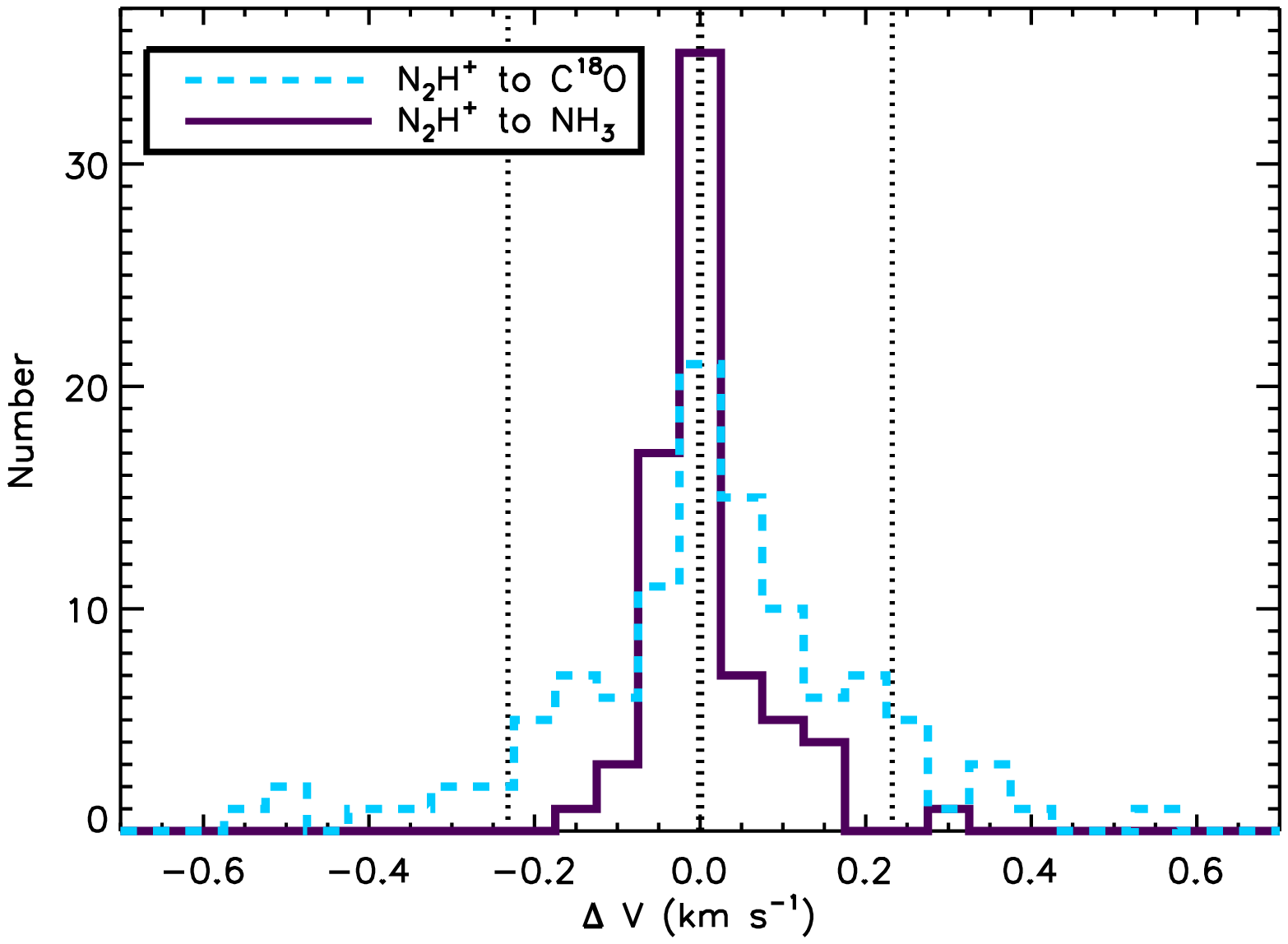}{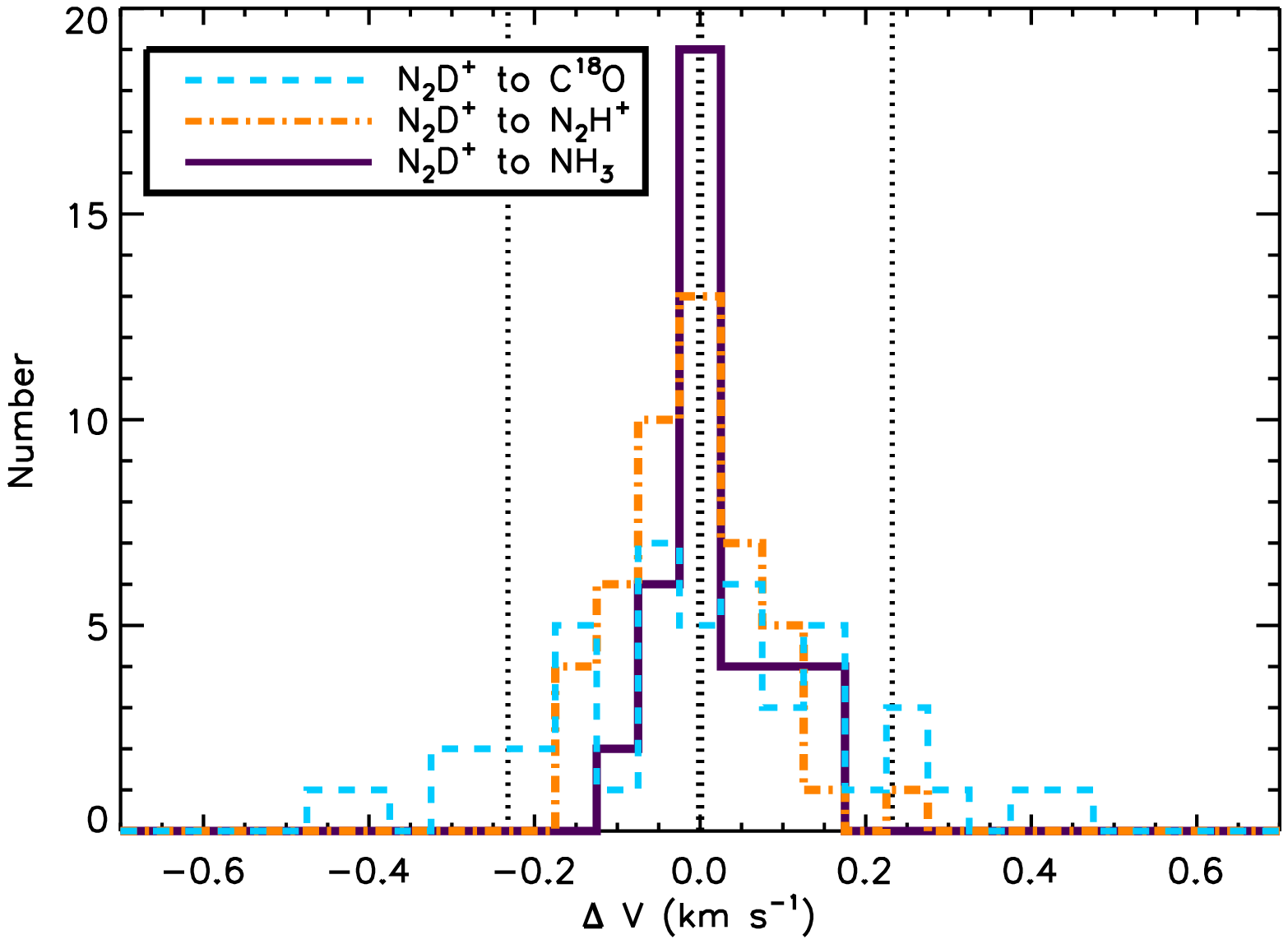}
\caption{Left: The distribution of centroid velocity differences for \dia\ to \amm\ and \dia\ to \co.  The vertical dotted lines denote the approximate sound speed in the region (assuming T=15~K), and 0~km~s$^{-1}$. Right: The distribution of centroid velocity differences for \ddia\ to \dia, \ddia\ to \amm, and \ddia\ to \co.  The vertical dotted lines denote the approximate sound speed in the region (assuming T=15~K), and 0~km~s$^{-1}$. \label{fig_vdiffs_n2dp}}
\end{figure*}

The centroid velocity of a core is an easily measured property. As described above, even with relatively low velocity resolution, we are able to determine the \ddia\, line centroid with good accuracy.  Previous work has shown that the velocity centroids measured for denser core gas, as traced by, e.g., \dia, tend to be similar to those measured for the surrounding, lower density envelope of material, as traced by, e.g., \co\ \citep{walsh04,walsh07,kirk07}.  For cores in the Perseus molecular cloud, the differences in centroid velocities between \dia\ and \co\ are small, typically less than the sound speed, but are often non-zero \citep[rms absolute velocity difference $\sim 0.17$~km~s$^{-1}$,][]{kirk07}. A much tighter relationship has been found for the centroid velocities of \dia\ and \amm\ in the Perseus cores, where the \amm\ data was taken from \citet{rosolowsky08}.  The mean absolute offset of \dia\ to \amm\ centroid velocities is 0.07\,\kms\, \citep{johnstone10}.  This measurement suggests that the \dia\ and \amm\ occupy a similar volume within the dense core.

Here, we compare the centroid velocities measured using the \ddia\, 3-2 transition to the various other line centroids already measured for Perseus dense cores.  Since \ddia\ is expected to trace a higher density of gas than even \dia, it could be expected to show an equal or larger typical velocity difference to the lower density gas tracers than \dia. The left panel of Figure~\ref{fig_vdiffs_n2dp} summarizes the previously observed velocity differences between \dia, \amm, and CO in the Perseus cores described in \citet{kirk07}, here including only cores which are part of this survey. The dashed blue line shows the velocity difference in \dia\ and \co\ observed in the \citet{kirk07} IRAM pointed survey, while the purple line shows the velocity difference between \dia\ and \amm. The right panel of Figure \ref{fig_vdiffs_n2dp} shows the comparable velocity differences found using \ddia. Similar to the trend seen in \dia, much larger centroid velocity differences are seen between \ddia\ and \co\ than \ddia\ and \amm. The velocity differences between \ddia\ and \dia\ appear to be slightly larger than for \ddia\ to \amm, although statistical tests (see below) show that there is no significant difference between them.  \showrev{Although the C$^{18}$O data were taken with significantly better angular resolution than the \dia\, and \ddia\, observations, the emission traces the outer regions of the core due to the effects of depletion and the lower critical density of the C$^{18}$O transition. C$^{18}$O observations made with a larger beam would include more emission from lower density material, and would likely show larger velocity differences between the tracers. }

We next compare the velocity difference distributions for \dia\, versus \amm\, and C$^{18}$O for all of the dense cores detected in \dia\, versus the subset observed and detected in \ddia\,, using a two- sided Kolmogorov-Smirnov (KS2) test \citep[eg,][]{conover99}. The KS2 test shows the likelihood that two distributions are drawn from the same parent population. The KS2 probabilities were found to be roughly 90\% or above, indicating that the cores in which \ddia\, was detected are representative, in terms of centroid velocity differences, of the full \dia\, sample of cores.  Furthermore, we compare the distribution of velocity differences seen between \ddia\, or \dia\, and \amm\, or C$^{18}$O, and found that the KS2 probabilities are sufficiently large (roughly 50\% or above) to suggest that \ddia\, and \dia\, behave similarly with respect to other gas tracers.

We also search for trends in the velocity difference between \ddia\ and \dia\ in the protostellar and starless core populations, as well as in the clustered and isolated cores. A comparison between the velocity differences for starless and protostellar cores shows that the former has a slightly smaller dispersion [$\sigma (\Delta V) = 0.071$\,\kms\, versus 0.091\,\kms]; however, a KS2 test suggests that these differences are not statistically significant.  Similarly, splitting the cores into those which are found in more isolated conditions and those found in more clustered environments shows that the former tend to have smaller velocity differences [$\sigma (\Delta V) = 0.067$\,\kms\, versus 0.097\,\kms], but again, a KS2 test shows these differences are not statistically significant. 

\subsection{\ddia\, column density and the \dia\, deuterium fractionation in Perseus cores}
\label{sec:rd}

Both the \ddia\, 2-1 and 3-2 transitions contain hyperfine structure, which allows independent calculation of the excitation temperature, $T_{ex}$, and the opacity, $\tau$, of the emitting gas. In practice, however, a high signal-to-noise ratio is needed in the satellite components to calculate $T_{ex}$ and $\tau$ with precision. \showrev{No \ddia\, 3-2 spectra showed significant hyperfine structure above the rms noise level. While hyperfine structure in the \ddia\, 2-1 line is seen toward multiple sources, our hyperfine line fits are unable to constrain well the \ddia\, 2-1 line opacity and the resulting uncertainties on $T_{ex}$ and $\tau$ are large (see Table \ref{tab:sourceDetections}).} In contrast, the observed cores were selected to be bright in \dia\, 1-0 emission, and consequently \citet{kirk07} were able to fit well the \dia\, hyperfine structure. The \dia\, excitation temperature can be determined from their fit results. 

\showrev{We expect that \dia\, and \ddia\, trace similar physical conditions given their similar kinematics in the Perseus cores (see \S \ref{sec:kin}). While we are unable to directly compare the excitation of the lines in our cores, previous studies have shown that the excitation of the two species are similar \citep{gerin01,caselli02}. We can then test our assumption that the \ddia\, lines are optically thin by performing a hyperfine fit to the line structure, setting $T_{ex}$ equal to that derived from the \dia\, 1-0 analysis. For both \ddia\, transitions, the total optical depths thus determined are universally $< 1$. We are thus confident that we can assume the \ddia\, emission is optically thin in the following analysis.}


\showrev{In the optically thin regime, we calculate the \ddia\, column density, $N(\mbox{\ddia})$, from the \ddia\, integrated intensity $W$ following the formula for optically thin emission from \citet{caselli02}:}

\begin{align}
N(\mbox{\ddia}) &= \frac{8 \pi W}{\lambda^3 A} \frac{g_l}{g_u} \frac{1}{J_\nu(T_{ex}) - J_\nu(T_{bg})} \nonumber \\
&\times \frac{1}{1 - \mbox{exp}(-h\nu / kT_{ex})} \frac {Q_{rot}}{g_l \mbox{exp}(-E_l / k T_{ex})}
\end{align}

\showrev{Here, $\lambda$ is the wavelength of the observed transition, $A$ is the Einstein coefficient ($A = 1.97 \times 10^{-4}$ and $7.14 \times 10^{-4}$ for \ddia\, 2-1 and 3-2, respectively), $g_l$ and $g_u$ are the statistical weight of the lower and upper levels, $h$ is Planck constant, $T_{ex}$ is the excitation temperature, $Q_{rot}$ is the partition function, $E_l$ is the energy of the lower level, and $k$ is the Boltzmann constant. $J_\nu(T_{ex})$ and $J_\nu(T_{bg})$ are the equivalent Rayleigh-Jeans excitation and background temperatures. For linear molecules like \ddia, $E_l = J_l ( J_l + 1) h B$ and $Q_{rot} = \sum_{l=0}^{\infty} g_l \exp(-E_l / (k T_{ex}))$, where $J_l$ is the rotational quantum number of the lower level, and the rotational constant $B = 38554 \times 10^6$\,Hz.} For each core, we use the \ddia\, 3-2 integrated intensity and set the excitation temperature, $T_{ex}$ to that derived from the \dia\, 1-0 analysis. For sources where \ddia\, 3-2 was not detected, we instead calculate the \ddia\, column density from the \ddia\, 2-1 integrated intensity. Where neither \ddia\, line was detected, we determine an upper limit to the integrated intensity based on the rms noise of the spectra, $\sigma_W$, following $\sigma_W = \Delta v_{res} \sigma_{rms} N_{chan}^{1/2}$.  Here, $\Delta v_{res}$ is the spectral resolution, $\sigma_{rms}$ is the 1$-\sigma$ noise value, and $N_{chan}$ is the number of channels over which the line integrated intensity was determined. The \ddia\, column density upper limit is then calculated using the \dia\, 1-0 $T_{ex}$ for the source and setting $W = 3 \times \sigma_W$ for each of the \ddia\, 2-1 and \ddia\, 3-2 transitions, and selecting the resulting lowest limit. 

\begin{figure*}
\plotone{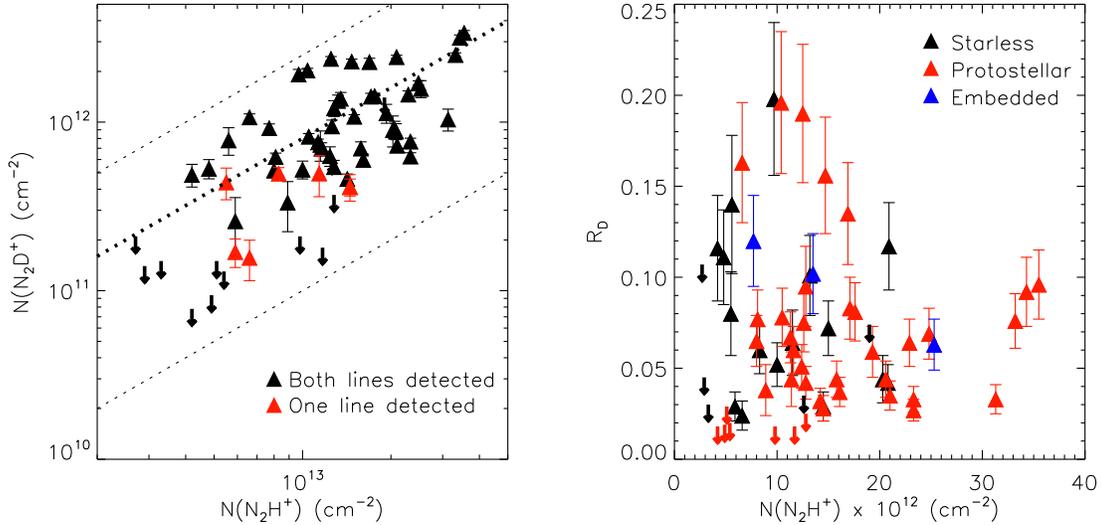}
\caption{Left: $N(\mbox{\ddia})$ vs. $N(\mbox{\dia})$ for the Perseus cores. Cores with both \ddia\, 3-2 and \ddia\, 2-1 detections are shown by solid black triangles, while cores with a single detection in either \ddia\, 3-2 or \ddia\, 2-1 are plotted with open black triangles. Arrows show the $N(\mbox{\ddia})$ upper limits for cores with non-detections in both transitions. The dotted lines represent constant \dia\, deuterium fraction levels, $R_D = 0.01$, the mean $R_D = 0.08$ for the sample, and $R_D = 0.25$. Right:  $R_D$ vs. $N(N\mbox{\dia})$ for the Perseus cores. Starless and protostellar cores are shown by black and grey (colored red in the online version) triangles, respectively, while deeply embedded sources are shown as open triangles (colored blue in the online version). \label{fig:NvsN}}
\end{figure*}

The deuterium fraction in a species, $R_D$, is defined as the ratio of the column density of the deuterated molecule with that of its hydrogen-bearing counterpart, where $R_D = N(\mbox{\ddia})\,/\,N(\mbox{\dia})$ for \dia. For sources which show no evidence of multiple velocity components along the line of sight, we calculate $N(\mbox{\dia})$ following \citet{caselli02} for optically thick transitions, using the hyperfine fit parameters in \citet{kirk07}. Two velocity components are visible in the \dia\, spectra for nine cores, but we are unable to separate velocity components in the \ddia\, spectra. To provide a more accurate estimate of the \dia\, deuterium fraction for these objects, we thus calculate the \dia\, column density using the total integrated intensity of the \dia\, line presented in \citet{kirk07}, following Equation A12 in \citet{johnstone10} and assuming optically thin emission. Where targets overlap with \citeauthor{johnstone10}, our \dia\, column densities agree within uncertainties. \showrev{The uncertainty in $R_D$ is propagated from the uncertainty in both $N(\mbox{\ddia})$ and $N(\mbox{\dia})$ \citep[taken to be 20\,\%;][]{johnstone10}. }

The resulting deuterium fractionation of \dia\, in each core is listed in Table \ref{tab:deutFrac}. We find a mean $R_D = 0.08$ for all cores, with a standard deviation of $0.04$. We plot $N(\mbox{\ddia})$ against $N(\mbox{\dia})$ in Figure \ref{fig:NvsN} (left), overlaid with lines of constant $R_D = 0.01, 0.08$, and 0.25. Cores in which one or both \ddia\, lines were detected are shown separately, and upper limits are shown for those cores where neither line was detected. Single-line detections are more likely at lower \ddia\, column densities, but span a range in \dia\, column density, and hence a range in $R_D$. 

In Figure \ref{fig:NvsN} (right), we plot $R_D$ as a function of $N(\mbox{\dia})$. No separation is seen in $R_D$ between protostellar and starless sources. Intriguingly, the highest $R_D$ values are found toward cores with relatively low \dia\, column densities compared with the entire sample. A similar result was found by \citet{fontani11} in their \dia\,/\,\ddia\, survey of massive star forming regions, which they suggest may be due to a decrease in the deuterium fractionation, while the \dia\, abundance continues to increase, in the most embedded protostellar phase. Figure \ref{fig:NvsN} (right) shows that on average, the protostellar targets in Perseus do have greater \dia\, column densities than the starless cores. The high $R_D$, low $N(\mbox{\dia})$ points include both starless and protostellar cores. We will show in \S\ref{sec:disc}, however, that these protostellar sources are consistent with being young. 

\begin{figure}
\epsscale{2.3}
\plotone{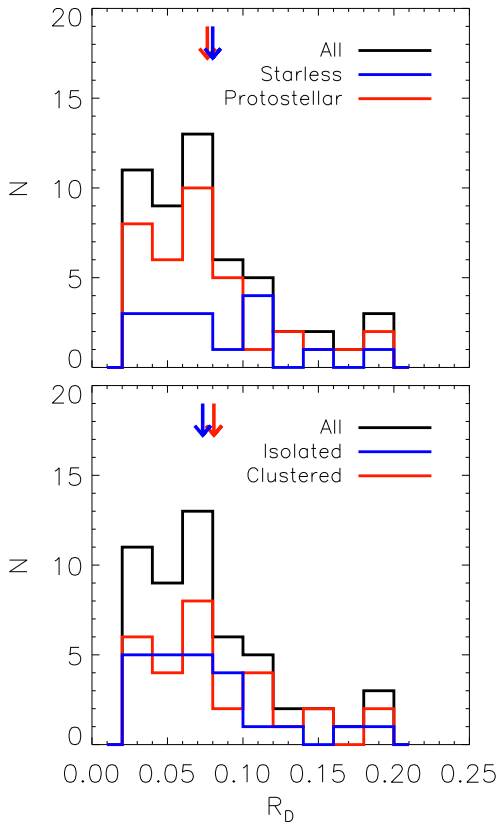}
\caption{Top:  Distribution of $R_D$ values found for Perseus cores. The total distribution is shown in black; blue and red histograms show the distribution for starless and protostellar cores, respectively. Mean values are shown with arrows. Bottom: Total $R_D$ distribution as in (top), with blue and red histograms showing the distribution for isolated and clustered cores, respectively. Mean values are shown with arrows. \label{fig:rdHist}}
\epsscale{1}
\end{figure}

We find that there is little difference in the deuterium fraction $R_D$ values between the starless and protostellar core subsamples, nor between the isolated and clustered core subsamples. All subsets have similar mean values and standard deviations to that of the overall sample. This can be seen clearly in Figure \ref{fig:rdHist}, where we show histograms of the $R_D$ values for the Perseus cores, separated by subsample. Most of the Perseus cores have deuterium fractionation values $R_D \lesssim 0.1$; however a small number of cores form a tail in the distribution to higher $R_D$ values. As the Figure shows, cores with higher deuterium fractionation are not confined to the starless or protostellar population, nor are they found only in clustered or distributed star forming environments. 

\begin{figure*}
\begin{center}
\epsscale{1.7}
\plotone{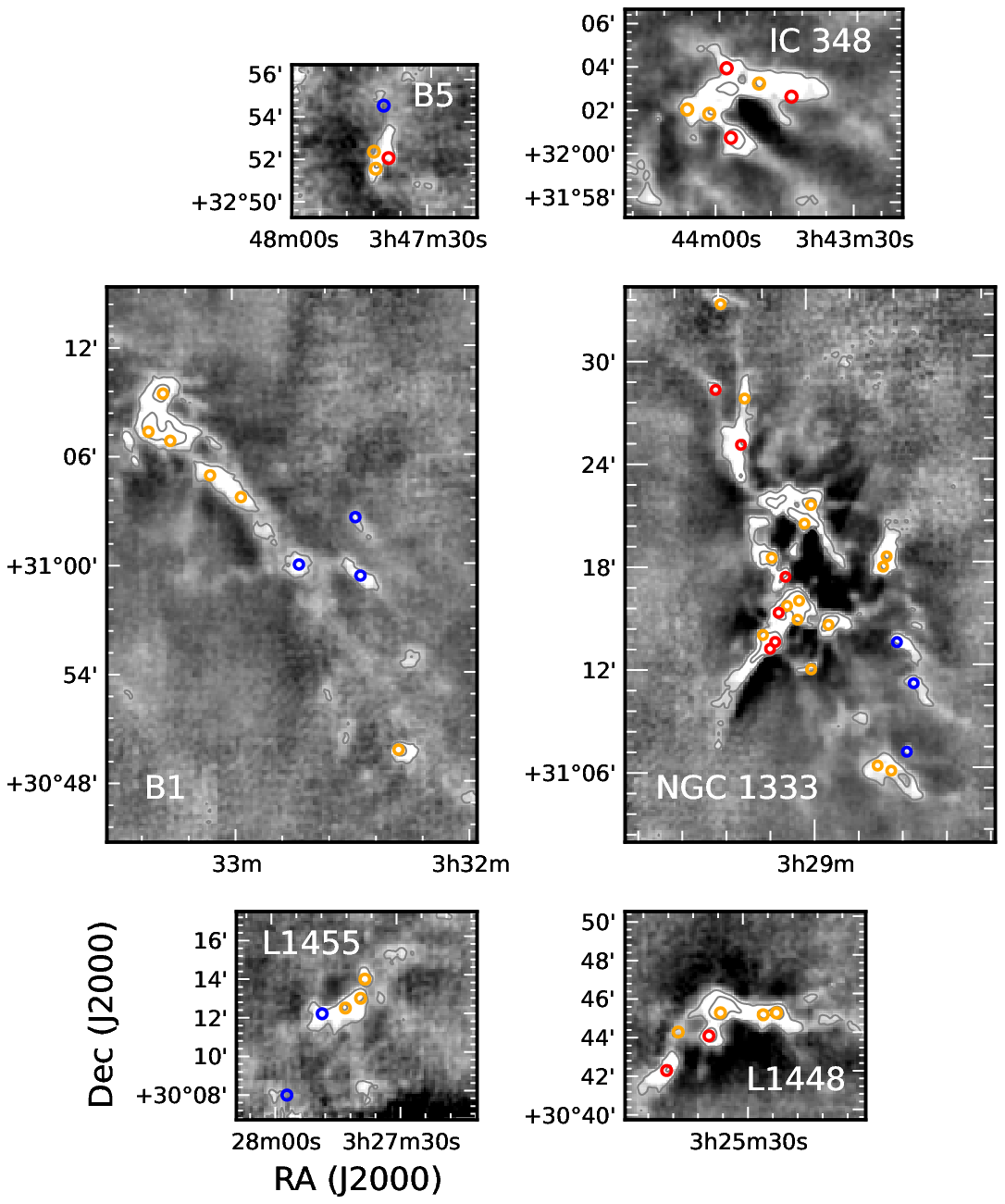}
\caption{850\,\micron\, continuum emission toward well-known star-forming regions within the Perseus molecular cloud taken with SCUBA on the JCMT \citep{kirk06}. In Perseus, the regions shown run approximately east to west from top left to bottom right. To better show the cloud structure, the greyscale has been truncated at 0.2\,Jy\,beam$^{-1}$, while peak values reach $\sim 8$\,Jy\,beam$^{-1}$ toward several regions in NGC 1333. Grey contours show flux densities of 0.1\,Jy\,beam$^{-1}$ and 0.5\,Jy\,beam$^{-1}$. Overlaid are circles (30\arcsec\, diameter) showing the positions of 54 of 64 cores observed in this study, with colours indicating their derived \dia\, deuterium fraction, $R_D$. Cores with $R_D > 0.1$ are shown in red, $R_D < 0.1$ are shown in orange, and \ddia\, non-detections are shown in blue.  The remaining ten cores are located outside these well-known regions: one shows $R_D > 0.1$, six show $R_D < 0.1$, and three do not have \ddia\, detections.
\label{fig:rd}}
\epsscale{1}
\end{center}
\end{figure*}

To further investigate the properties of the high-$R_D$ cores, we show in Figure \ref{fig:rd} the distribution of $R_D$ values across the Perseus molecular cloud, colored to show cores with $R_D \geq 0.1$, $R_D < 0.1$, and \ddia\, non-detections. Most of the well-known groupings of star forming cores (e.g., B5, IC348, NGC 1333 and L1448) contain at least one core with $R_D \geq 0.1$, as well as a few moderately deuterated cores and \ddia\, non-detections, and we see no systematic trend of the \dia\, deuterium fraction across the cloud as a whole. Five cores have $R_D > 0.15$. Comparing the physical properties of these cores with the rest of the sample, we find that the high $R_D$ cores are characterized by greater 850\,\micron\, continuum flux and core concentration, larger column densities of H$_2$, as well as higher \amm\, and lower \dia\, column densities (as seen in Figure \ref{fig:NvsN} (right). 

\showrev{Similar results have been found previously in Perseus. For example, toward three cores in Perseus, \citet{roueff05} detected triply-deuterated ammonia, ND$_3$, with an abundance of 0.1\,\% relative to \amm, suggesting the abundance of singly-deuterated \amm\, is $\sim 10$\,\%.} The average $R_D$ values found here also agree well with other studies of low-mass cores. \citeauthor{crapsi05} find a mean $R_D = 0.1$ for their core sample, with values ranging from 0.03 to 0.44. Similarly, other studies of low mass starless and protostellar cores find $R_D$ values within the range of a few hundredths to a few tenths \citep{daniel07,emprechtinger09,roberts07,friesen10b,gerin01,miettinen09,pagani09}. Interestingly, no core in Perseus shows \dia\, deuterium fractionation levels above $\sim 0.25$, despite our core sample spanning mean and central densities up to $n \sim 10^6$\,\cc. Regions of high deuteration exist in Perseus, but are not apparent in our analysis. In NGC 1333, for example, \citet{vandertak02} detected triply-deuterated \amm\, toward IRAS 4A in a 25\arcsec\, beam, where we find $R_D = 0.16$. 

Ten of the sources analysed by \citeauthor{emprechtinger09} overlap with protostellar cores in our Perseus sample. A direct comparison of the deuterium fractionation values for the overlapping targets shows that their results are greater, on average, than those reported here, with varying discrepancies that range from a factor of 0.8 to 3. \citeauthor{emprechtinger09} derive both $N(\mbox{\dia})$ and $N(\mbox{\ddia})$ from observations taken with the IRAM 30\,m telescope based on the \dia\, $J=1-0$ and \ddia\, $J=2-1$ transitions, respectively. While the \dia\, observations (and consequently the derived column densities) thus match ours in transition and beam size, the angular resolution of the \ddia\, 2-1 observations (16\arcsec) is significantly better than both our \ddia\, 2-1 (41\arcsec) and 3-2 (30\arcsec) observations. In all but one overlapping source, \citeauthor{emprechtinger09} report greater integrated \ddia\, 2-1 line intensities than found in our observations. This suggests that the filling factor of the \ddia\, emission, here assumed to be unity, is $\lesssim 1$ for some sources. We discuss the filling factor assumption and its effect on our analysis further in \S \ref{sec:ff}. 

\subsubsection{Trends in the deuterium fractionation}
\label{sec:trends}

\begin{figure*}[t]
\plotone{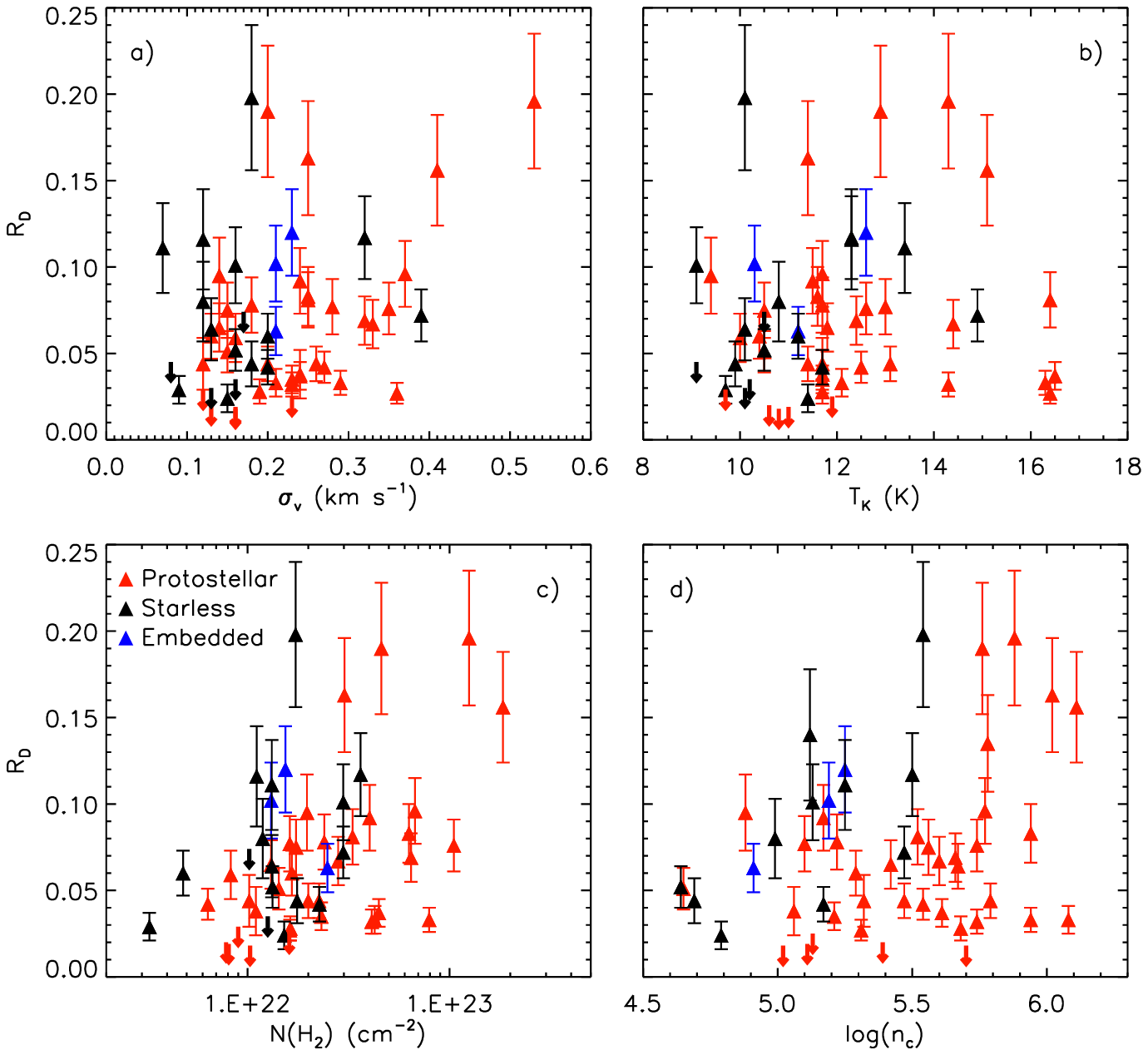}
\caption{Trends in the \dia\, deuterium fractionation, $R_D$, with \dia\, line width ($\sigma_v$), gas temperature ($T_K$), H$_2$ column density ($N(\mbox{H$_2$})$), and the log of the central density, $n_c$. The gas temperature is calculated from analysis of \amm\, (1,1) and (2,2) inversion transition emission \citep{rosolowsky08}. $N(\mbox{H$_2$})$ is calculated from 850\,\micron\, continuum measurements \citep{kirk06}, recalculated here using the \amm-derived gas temperature.  The central density is derived from a Bonnor-Ebert sphere analysis of the 850\,\micron\, continuum emission toward the cores \citep{kirk06}. In all plots, starless and protostellar cores are shown by black and grey (colored red in the online version) triangles, respectively, while deeply embedded sources are shown as open triangles (colored blue in the online version).  \label{fig:rdTrends}}
\end{figure*}

We plot in Figure \ref{fig:rdTrends} the \dia\, deuterium fraction against several core parameters thought to be indicators of core evolutionary stage: the \dia\, 1-0 velocity dispersion, $\sigma_v$, the kinetic gas temperature, $T_K$, and the column density of H$_2$, $N(\mbox{H$_2$})$. Where targets coincide with a SCUBA core in the \citet{kirk06} catalogue (51/64 sources), we also plot $R_D$ against the core central density, $n_c$, from a Bonnor-Ebert (BE) fit to the data. For all parameters plotted, we test for significant correlations using Spearman's $\rho$ rank correlation test \citep{conover99}. 

The H$_2$ column density, $N(\mbox{H$_2$})$, is calculated from SCUBA 850\,\micron\, continuum emission \citep[$S_\nu$; described in ][]{kirk06}, following the standard modified blackbody equation $N(\mbox{H}_2) = S_\nu\, /\, (\Omega_m \mu m_H \kappa_\nu B_\nu T_d)$. Here, $\mu = 2.33$ is the mean molecular weight, $m_H$ is the mass of hydrogen, $\kappa_\nu = 0.02$\,cm$^{2}$\,g$^{-1}$ is the dust opacity per unit mass at 850\,\micron, and $B_\nu (T_d)$ is the Planck function at the dust temperature $T_d$.  We sum flux over a 30\arcsec\, beam ($\Omega_m$) for each target. The dust opacity is taken from the dust models of \citet{ossenkopf} for agglomerated grains with ice mantles (OH5). This value is consistent with observational results in low mass star forming cores \citep{shirley02}. We set the dust temperature equal to the \amm-derived gas temperature where cores overlap with \citet{rosolowsky08}. Elsewhere, we set $T_d = 11$\,K, the mean of the Perseus sample. Our $N(\mbox{H$_2$})$ values thus differ slightly from those reported for overlapping cores by \citet{johnstone10}, who set $T_d = 11$\,K (the mean value for all cores in their study) for all cores. \showrev{There are several sources of uncertainty in the $N(\mbox{H}_2)$ values. First, the calibration uncertainty in the continuum flux values is $\sim 20$\%. Second, the dust properties are not known with high precision. For example, estimates of $\kappa_{850}$ can vary by $\sim 3$ \citep{shirley00}, while $T_d$ may differ from $T_K$. At the densities typical of the cores studied here ($n \gtrsim 10^5$\,\cc), however, we expect that the dust and gas are coupled \citep{goldsmith01}. The BE analysis independently derived core temperatures of $T = 10-19$\,K for the Perseus cores, in general agreement with the derived $T_K$ values we use here. In addition, while $\kappa_{850}$ may not be known to high precision, we do not expect large variations in $\kappa_{850}$ between the cores given their similar temperatures and densities. We estimate the H$_2$ column densities have uncertainties of factors of a few.} 

We find no trend in the deuterium fractionation of the cores as a function of \dia\, line width (Figure \ref{fig:rdTrends}a. Similar results were found by \citet{crapsi05}, and by \citet{roberts07} toward a sample of low- and high-mass protostellar cores. Toward cores in massive star forming regions, \citet{fontani11} find a faint anticorrelation between $R_D$ and \dia\, line width ($J=3-2$ transition). \citet{chen11} also find an anticorrelation between $R_D$ and the \ddia\, line width ($J = 3-2$ transition) toward massive protostellar/cluster cores. Anticorrelation between deuterium fractionation and line width may be expected if the turbulence typical of the interstellar medium has dissipated within the most evolved dense cores, resulting in gas motions that are nearly purely thermal. Dense gas within gravitationally unstable cores may also, however, have significant infall motions ($\sim 0.1$\,\kms) traced by \dia\, \citep{schnee07,difran01}, which would add a broader component to the observed line widths and may confuse any trend with deuterium fraction.

We additionally find no trend in the deuterium fractionation of the cores as a function of gas temperature, $T_K$ (Figure \ref{fig:rdTrends}b). Multiple studies have found significant anticorrelation between temperature and deuterium fractionation, including \citet{chen11,emprechtinger09,fontani11}. This anticorrelation is expected for several reasons. First, the reaction that forms H$_2$D$^+$ from H$_3^+$ is exothermic, and at low temperatures can only proceed in the forward direction. This results in a reservoir of \hd\, to react with molecules in the gas phase, and produce deuterated species. At temperatures greater than $\sim 25$\,K, \hd\, is able to react with H$_2$ and the \hd\, abundance decreases \citep{roberts07}. Second, at low temperatures ($\sim 20-30$\,K) CO becomes depleted from the gas phase through freeze-out onto dust grains, removing a destruction pathway for \hd\, (and \ddia) and thus allowing enhanced abundances of deuterated species. When a protostar forms and begins warming the nearby gas, CO can desorb from the dust grain, returning to the gas phase and again reacting with \hd\, (and \ddia) to suppress deuterium fractionation in molecular species. In the studies already mentioned, the decrease in deuterium fractionation is largely seen at temperatures $\gtrsim 16 - 25$\,K. Because all the Perseus cores studied here have gas temperatures $T_K \lesssim 17$\,K, it is reasonable that Figure \ref{fig:rdTrends}b shows no strong correlation between deuterium fractionation and temperature. 

We note, however, that the measured gas temperatures for our targets which overlap with those of \citet{emprechtinger09} are significantly colder (often by factors of 2 or more) than the dust temperatures the authors determined through spectral energy distribution (SED) fitting, and where the authors find a strong anti-correlation between $R_D$ and $T_{d}$. At the densities typical of the cores studied here ($n \gtrsim 10^5$\,\cc), we expect that the dust and gas are coupled \citep{goldsmith01}. Since warm dust radiates more efficiently, however, it is possible that the average derived dust temperature of a core with a warm central component (i.e., due to heating from a young protostar) will be greater than that probed by \amm\, observations. Compared with our results at larger angular resolution, this suggests that the average deuterium fraction of dense gas-tracing molecules in the protostellar cores is dominated by the physical conditions close to the protostar. A similar result was found previously by \citet{friesen10b} toward the Ophiuchus B core, where the deuterium fraction in \dia\, and \hd\, abundance both decrease with increasing proximity to protostars embedded within the core.

The large discrepancy between $T_d$ and $T_K$ could also be due to the difficulty in fitting dust SEDs with few data points, however, where anticorrelations between $T_d$ and the dust opacity power law index, $\beta$, can be mimicked by noise in the data points \citep{shetty09a,shetty09b}. While \citeauthor{emprechtinger09} model the SEDs of their sources using $\beta = 2$, consistent with the OH5 dust opacity model described above including a strong gradient in $T_d$ in their models, they then determine a single dust temperature $T_d$ from an isothermal fit to the data, assuming $\beta = 1$. The higher dust temperatures determined through this analysis may better reflect the physical conditions of the dust close to the protostar than the $T_K$ measurements. Since our observations provide a beam-averaged measure of the deuterium fraction, however, we conclude that the $T_K$ measurements provide the best estimate of the core temperature for our analysis.

We find in Figure \ref{fig:rdTrends}c that the \dia\, deuterium fraction tends to increase with the column density of H$_2$, $N(\mbox{H$_2$})$. Although there is substantial scatter, a statistical test shows the correlation is significant (Spearman parameter $\rho = 0.29$, with a significance of 0.04, where low values indicate greater significance). Considering the starless and protostellar populations separately, we find that the deuterium fractionation in the protostellar population is correlated with $N(\mbox{H$_2$})$ ($\rho = 0.36$, significance 0.04 omitting the embedded sources, and $\rho = 0.27$, significance 0.12 including the embedded sources). There is no significant correlation in the starless population. Similarly, a general increase in $R_D$ with central density $n_c$ is seen in Figure \ref{fig:rdTrends}d when considering  the entire core sample, where $\rho = 0.23$ with a significance of 0.1. In this case, however, the correlation is stronger when analysing the starless cores separately ($\rho = 0.55$, significance of 0.02).  


\subsubsection{Deuterium fractionation and CO depletion}
\label{sec:co}

Models of dense gas chemistry show that the depletion of CO from the gas phase increases the rate of deuterium fractionation reactions in the gas phase \citep{dalgarno}. At low temperatures, the deuterium fractionation rate is strongly dependent on the CO depletion \citep{caselli02,caselli08}. Recent studies have indeed found strong correlations between the deuterium fractionation of \dia\, (and other molecular species) and the amount of CO depletion, $f_D$, in a star-forming core \citep{caselli02,crapsi04,crapsi05,caselli08,emprechtinger09,chen11}. Here we determine the beam-averaged CO depletion toward the Perseus cores using the C$^{18}$O data presented in \citet{kirk07}. The CO depletion is then $f_D = x_{can} (\mbox{C$^{18}$O}) / x(\mbox{C$^{18}$O})$, where $x_{can}(\mbox{C$^{18}$O})$ is the canonical abundance of C$^{18}$O relative to H$_2$ in the ISM. To best compare our results with \citet{crapsi05}, we set $x_{can} = 1.7 \times 10^{-7}$ \citep{frerking82}, but note that other values of $x_{can}$ exist in the literature \citep[e.g.,][]{lacy94}. We discuss further the effects of using the \citeauthor{frerking82} value in \S\ref{sec:chem}. 

To calculate $N(\mbox{C$^{18}$O})$, we assume that the C$^{18}$O emission is optically thin, and set $T_{ex} = T_K$ for all sources. We then derive $N(\mbox{C$^{18}$O})$ in the optically thin case using the line integrated intensity, following \citet{caselli02}, and determine $f_D$ for each source using the previously derived $N(\mbox{H$_2$})$ (\S \ref{sec:trends}). We expect the assumption that $T_{ex} = T_K$ to be reasonable for the Perseus cores, given that C$^{18}$O traces lower density gas than \amm. At lower densities, the gas-dust coupling is less efficient, and the gas kinetic temperature may be elevated above the dust temperature by photoelectric heating. In this case, the C$^{18}$O kinetic temperature, $T_{K, CO}$, may be somewhat higher than for \amm. Since the C$^{18}$O is likely subthermally populated, however, $T_{ex} < T_{K,CO}$. If the C$^{18}$O emission has moderate opacity, $N(\mbox{C$^{18}$O})$ would be underestimated (and $f_D$ would be overestimated) by $\sim 30$\% for $\tau = 0.5$, and $\sim 60$\% for $\tau = 1$. \showrev{The uncertainty in the CO depletion also depends linearly on the uncertainty in $N(\mbox{H}_2)$, which may be relatively large as described above. For example, if we have overestimated the dust emissivity at 850\,\micron\, for the Perseus molecular cloud, our derived $N(\mbox{H}_2)$ values will be systematically low, as will our derived CO depletion factors. Variations in the dust properties between the cores will introduce scatter in the results. }Two cores lie off the area mapped by SCUBA, allowing no measurement of $N(\mbox{H$_2$})$, and consequently $f_D$. 

On average, Perseus cores are characterized by low values of $f_D$. CO depletion factors $f_D > 5$ are found only toward three protostellar cores, including NGC 1333 IRAS 4A and IRAS 4B (our sources 106 and 103), where $f_D = 8$ and $f_D = 19$, respectively. Both IRAS 4A and IRAS 4B have been previously shown to have high depletion of CO \citep{blake95}, in agreement with our results. More recently, by modeling the observed emission from seven C$^{18}$O transitions, \citet{yildiz12} showed that CO abundance profiles toward both sources must include a factor of $> 10$ drop in abundance between $\sim 5 \times 10^2$\,AU and $\sim 10^4$\,AU (where $10^4$\,AU is $\sim 40$\arcsec\, at the distance of Perseus). As expected, IRAS 4A and 4B are also two of the Perseus cores that show the highest \dia\, deuterium fraction levels, with $R_D = 0.16$ and $R_D = 0.20$, respectively. 

All other cores have CO depletion factors $f_D \lesssim 5$, with a mean $f_D = 2.5$ for the sample. We additionally find $f_D < 1$ for 12 cores, meaning that these cores contain more C$^{18}$O relative to H$_2$ than expected from the canonical value. Seven of these over-abundant C$^{18}$O cores lie in the eastern part of Perseus. For example, cores in the B5 and IC348 regions have a mean $f_D = 1.0$ and 1.3, respectively, with small standard deviations of 0.5 and 0.7. Cores in NGC 1333, L1448, L1455 and B1 show greater average $f_D$ values, with a mean $f_D \sim 3-4$, and substantially larger variations in $f_D$ between cores in each region. The results are suggestive of a gradient in the C$^{18}$O depletion across Perseus, but we note that cores in B5 and IC348 together include only 13 of our sample of 64 cores. 

\begin{figure}
\epsscale{1.2}
\plotone{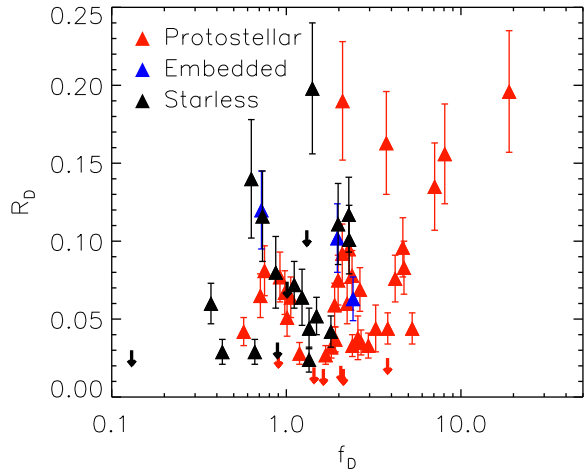}
\caption{The \dia\, deuterium fractionation for Perseus cores as a function of CO depletion, $f_D$, as measured through observations of C$^{18}$O 2-1 emission at 11\arcsec\, resolution. Starless and protostellar cores are shown by black and grey (colored red in the online version) triangles, respectively, while deeply embedded sources are shown as open triangles (colored blue in the online version). \label{fig:fd}} 
\epsscale{1}
\end{figure}

We plot in Figure \ref{fig:fd} the \dia\, deuterium fraction against the CO depletion factor for each core. In general, the Figure shows that $R_D$ tends to increase with $f_D$, although the trend is most significant for the protostellar sample, omitting the most embedded sources (Spearman's $\rho = 0.34$, with a significance of 0.05). On average, protostellar cores are also characterised by significantly greater CO depletion (mean $f_D = 3.2$) than starless cores (mean $f_D = 1.2$), with a $<1$\,\% KS2 probability that the distributions are drawn from the same sample. For ten cores in our sample, \citet{emprechtinger09} also calculated the CO depletion factor using observations of C$^{18}$O 1-0 (22\arcsec\, HPBW). They find $f_D$ values that are generally within factors of $\sim 2$ of the results presented here. Most (but not all) of our $f_D$ values are greater than those presented by \citeauthor{emprechtinger09} This is reasonable considering our C$^{18}$O observations have finer angular resolution, and are thus sampling emission closer to the core center, encompassing less of the lower density gas where we expect lower CO depletion. 

It is interesting that many of our cores show low $f_D$ values. Low CO depletion will strongly decrease the deuterium fractionation, and models by \citet{caselli08} predict $R_D < 0.3$ for $f_D < 10$, in agreement with our results. Trends between $R_D$ and $f_D$ found in previous studies often show a marked increase in $R_D$ at $f_D > 10$ \citep{crapsi05,emprechtinger09}. At low $f_D < 10$, \citet{chen11} find a weak increasing trend in $R_D$ with $f_D$ in their study of massive protostellar cores. Recently, \citet{ford11} show, through radiative transfer modeling of C$^{18}$O data toward eight starless cores in Taurus, that $f_D$ is often very poorly constrained, largely due to uncertainties in the canonical abundance of C$^{18}$O and in the core temperature. They find that it is difficult to distinguish between $f_D = 10$ and $f_D = 1000$ in their models. In Perseus, however, even if the absolute $f_D$ values are uncertain, the finding that Perseus cores have low CO depletion on average is likely robust\showrev{, with the caveat that uncertainties in the dust emissivity may have a systematic effect on our results}. We discuss the implications of the low depletion values in Perseus in more detail in \S\ref{sec:chem}. 

\begin{figure*}
\epsscale{1}
\plotone{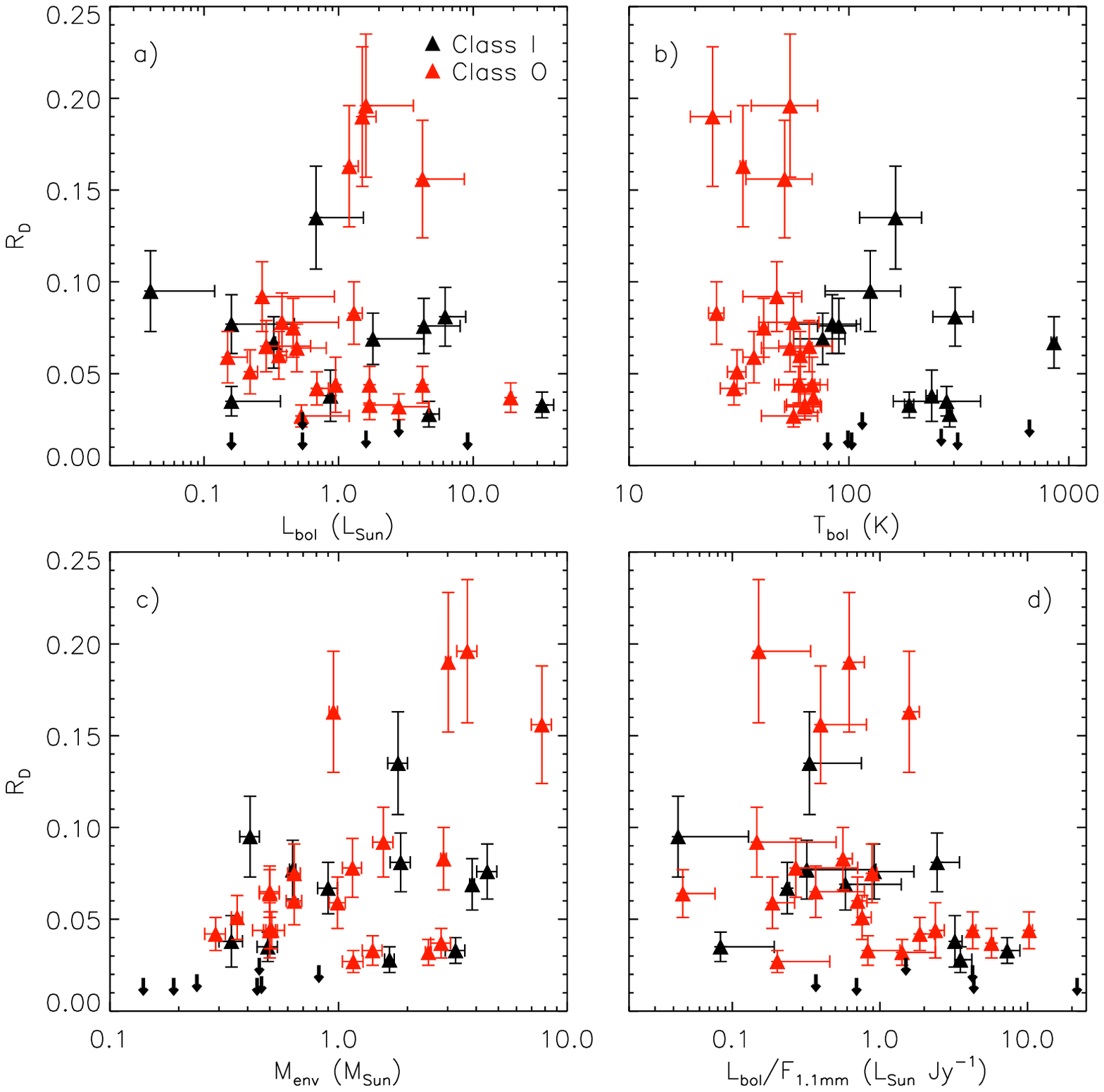}
\caption{The \dia\, deuterium fractionation of protostellar cores in Perseus against (a) $L_{bol}$, (b) $T_{bol}$, (c) the envelope mass $M_{env}$, and (d) the ratio of $L_{bol}$ to total 1.1\,mm flux. $L_{bol}$, $T_{bol}$ and $M_{env}$ values are taken from \citet{enoch09}, while the total 1.1\,mm flux is taken from \citet{enoch06}.  These catalogs do not include recently discovered very embedded sources. In all plots, Class 0 and Class I sources, as determined by \citeauthor{enoch09}, are shown grey (colored red in the online version) and black points, respectively. For clarity, we show one-sided x-axis error bars on the $L_{bol}$ plots. \label{fig:lbol}}
\epsscale{1}
\end{figure*}

\subsubsection{Deuterium fractionation and protostellar properties}

Finally, we examine the \dia\, deuterium fraction of the protostellar Perseus cores with respect to the properties of the embedded young stellar objects (YSOs). In their analysis of YSOs in Perseus, \citet{enoch09} determine the bolometric temperature ($T_{bol}$) and luminosity ($L_{bol}$), as well as envelope mass ($M_{env}$), based on spectral energy distribution fits to 2MASS, Spitzer IRAC and MIPS photometry, and Bolocam 1.1\,mm data ($F_{1.1\,mm}$). Younger sources are expected to have lower luminosities and temperatures, greater remaining envelope mass, and low ratios of $L_{bol}$ to submillimetre flux density. Of the 43 protostellar cores surveyed here, 39 have associated $T_{bol}$, $L_{bol}$, $M_{env}$ and $F_{1.1\,mm}$ measurements, of which we detect 32 in \ddia\, emission. Within this sample, 21 are identified as Class 0 sources and 18 as Class I, where Class 0 sources are defined as sources with $T_{bol} < 70$ (\citeauthor{enoch09}). We show in Figure \ref{fig:lbol} the deuterium fraction of these 39 sources as a function of $L_{bol}$, $T_{bol}$, $M_{env}$ and $L_{bol}/F_{1.1\,mm}$. Note that the three deeply embedded protostellar sources, which were not identified in the infrared surveys, have estimated internal luminosity upper limits $L_{int} \lesssim 0.01\,L_\odot$ \citep[Per-Bolo 45 and Per-Bolo 58;][]{schnee12,enoch10}, or bolometric luminosity upper limit $L_{bol} < 0.1\,L_\odot$ \citep[L1448-IRS2E;][]{chen10}. 

While the mean $R_D$ for the Class 0 sources is slightly greater than that for the Class I sources ($R_D = 0.08$ and 0.07, respectively), the difference is not significant.  We find no significant trends in $R_D$ with protostellar $L_{bol}$, $M_{env}$, or $L_{bol}/F_{1.1\,mm}$, but do find an anti correlation between $R_D$ and $T_{bol}$ (Spearman $\rho = -0.25$, significance of 0.07). In addition, the sources with greater $R_D$ values are found toward sources with $M_{env} \gtrsim 0.8$\,M$_\odot$, despite the lack of a clear correlation between $R_D$ and $M_{env}$. Furthermore, all the protostellar sources in the \citeauthor{enoch09} sample that we did not detect in \ddia\, emission have higher $T_{bol}$ values, identifying them as older, Class I sources, while six of seven undetected sources have $M_{env} \lesssim 0.5\,M_\odot$ (see $R_D$ upper limits in Figure \ref{fig:lbol}c). Finally, sources with high $L_{bol} / F_{1.1\,mm}$ values tend to have the lowest $R_D$ measurements, although there is significant scatter at $L_{bol} / F_{1.1\,mm} \lesssim 2-3$\,L$_\odot$\,Jy$^{-1}$. Taken together, these plots show that the deuterium fraction in \dia\, in the protostellar Perseus cores is consistent with the expectation that $R_D$ will drop as the protostar warms its immediate environment (releasing CO from dust grains into the gas) and additionally loses envelope mass to accretion and outflow, and is in agreement with the findings of \citet{emprechtinger09}. 

\section{Discussion}
\label{sec:disc}


\subsection{Potential complications of a pointed survey}

\subsubsection{Excitation effects} 
\label{sec:excite}

In \S \ref{sec:rd}, we set the excitation temperature of the \ddia\, emission equal to that of the \dia\, emission for each core to determine $N(\mbox{\ddia})$. This assumption has been shown to be accurate in several dense cores through high sensitivity observations \citep{caselli02,gerin01}. For those sources in this study where we detect both \ddia\, transitions, we can instead simultaneously determine both the column density of \ddia, $N(\mbox{\ddia})$, and $T_{ex}$, under the assumptions that both lines are optically thin and emission fills the beams, and requiring the excitation temperatures and total \ddia\, column density derived from each transition to be equal. Uncertainties in the derived properties were calculated using the formal uncertainties in the measured integrated intensities in a Monte Carlo analysis. In this analysis, using the column density calculation for optically thin emission in \citet{caselli02}, the excitation temperature derived from the two species is solely dependent on the ratio of the integrated intensities, $W_{2} / W_{1}$, where 

\begin{align}
\frac{W_2}{W_1} = & \frac{g_{u,2} \lambda_2^3}{g_{u,1} \lambda_1^3}\frac{A_2}{A_1} \frac{J_{\nu_2}(T_{ex}) - J_{\nu_2}(T_{bg})}{J_{\nu_2}(T_{ex}) - J_{\nu_2}(T_{bg})} \times \nonumber \\
& \frac{1- \mbox{exp}(- h \nu_2 / k T_{ex})}{1-\mbox{exp}(- h \nu_1 / k T_{ex})} \frac{\mbox{exp}(-E_{l,2}/k T_{ex})}{\mbox{exp}(-E_{l,1}/k T_{ex})}  .
\label{eqn:wRat}
\end{align}

\noindent Here, $g_{u,1}$ and $g_{u,2}$ are the upper state statistical weights for the two transitions, $\lambda_1$ and $\lambda_2$ are the transition wavelengths, $A_1$ and $A_2$ are the Einstein A coefficients, $E_{l,1}$ and $E_{l,2}$ are the lower state energy levels. $J_{\nu_1}(T_{ex})$ and $J_{\nu_2}(T_{ex})$ are the equivalent Raleigh-Jeans excitation temperatures for the two transitions at excitation temperature $T_{ex}$; $J_{\nu_1}(T_{bg})$ and $J_{\nu_2}(T_{bg})$ are the same for the background temperature, $T_{bg} = 2.73$\,K. The derived \ddia\, column density depends both on $T_{ex}$ and the measured integrated intensities. 

The resulting \ddia\, $T_{ex}$ values are often, but not always, in agreement within uncertainties with those measured from \dia\, 1-0 observations \citep{johnstone10}. Further separation of these sources into protostellar and starless populations  reveals that the \ddia\, $T_{ex}$ values for protostellar sources are greater, with a mean $T_{ex} = 8.4$\,K and a standard deviation of 2.5\,K, compared with starless sources, where the mean $T_{ex} = 6.7$\,K with a standard deviation of 1.8\,K. There is little difference, however, in the \dia\, $T_{ex}$ between the starless (mean $T_{ex} = 5.9$\,K) and protostellar (mean $T_{ex} = 6.0$\,K) populations, with the \dia\, and \ddia\, excitation temperatures consistent within uncertainties for the starless cores. Greater \ddia\, excitation temperatures for protostellar cores, if real, suggest that the \ddia\, observations may trace somewhat warmer or higher density gas near the embedded protostars than \dia\, 1-0. \showrev{Alternatively, however, this increase in measured $T_{ex}$ may be caused instead by variation in the emission filling factor within the telescope beams for \ddia\, 2-1 and \ddia\, 3-2 emission. }

\subsubsection{Source sizes}
\label{sec:ff}

\showrev{The excitation analysis is strongly dependent on the assumption that the emission for both species fills the respective beams (the filling fraction for both transitions is unity). The FWHM beam widths are 41\arcsec\, and 30\arcsec\, for the \ddia\, 2-1 and \ddia\, 3-2 transitions, respectively. To test this, we  included a filling factor $f$ equal to the ratio of the beam areas between the \ddia\, 3-2 and \ddia\, 2-1 observations and determined again $T_{ex}$ and $N(\mbox{\ddia})$ following Equation \ref{eqn:wRat}. Essentially, this calculation assumes that the size of the emitting region is similar to the \ddia\, 3-2 beam FWHM. When incorporating the beam ratio in the analysis, we find the excitation temperature of the \ddia\, emission is within uncertainties of the \dia\, $T_{ex}$ for most sources. This agreement suggests that the size of the \ddia\, emitting region is indeed smaller than the \ddia\, 2-1 beam, and likely similar to the \ddia\, 3-2 beam.  }

\showrev{Furthermore, if we assume the excitation of the \ddia\, emission is equal to the \dia\, excitation, we can determine the relative filling fraction of the beam between the two \ddia\, transitions, although we cannot determine accurately the size of the emitting region if it is significantly smaller than 30\arcsec. The resulting \ddia\, source sizes range from $\sim 25$\arcsec\, to $\sim 40$\arcsec, with a mean of 28\arcsec. This is in broad agreement with the source sizes from the SCUBA 850\,\micron\, continuum measurements, although individual source sizes do not agree well. Previous studies have shown that the emitting region of \ddia\, can be smaller than the SCUBA source size \citep[see, e.g.,][]{caselli02b}. This result is expected given that \ddia\, should be emitted from regions of high density, whereas the dust continuum emission will have some contribution from lower density material (although submillimeter and millimeter continuum observation methods generally result in the loss of emission on large angular scales, removing some of the lower density component).  In addition, we are able to directly compare the observed integrated intensity of \ddia\, 2-1 toward seven protostellar sources in our sample which overlap with IRAM \ddia\, 2-1 observations by \citet{emprechtinger09}. The measured \ddia\, integrated intensities agreed for only one source (Per4N in Emprechtinger et al.;  target 97 in our catalogue). For the rest, the \citet{emprechtinger09} integrated intensities were significantly greater. The resulting filling fractions for the overlapping targets in \ddia\, 2-1 emission range between 0.23 and unity, or source sizes of 20\arcsec\, to at least 41\arcsec\, (the FWHM of our \ddia\, 2-1 observations).  }

\showrev{We find that the emitting region sizes derived from these tests are significantly larger for protostellar cores than for starless cores, and we additionally find a positive trend between source size and core concentration. This is in agreement with previous results that the more massive cores in Perseus tend to be protostellar \citep{hatchell08}, and that highly concentrated cores are also more likely to be protostellar \citep{walawender05}. In sum, protostellar cores are more massive, more concentrated, and show \ddia\, emission over larger areas, but on average the deuterium fractionation in N2H+ remains the same as in starless cores. }

\showrev{For most of the cores, however, we derive $R_D$ from the integrated \ddia\, 3-2 intensity. The effect of variable filling fraction is therefore lessened, where the smallest filling fraction of 0.23 calculated for \ddia\, 2-1 from the \citeauthor{emprechtinger09} results increases to 0.44 in the \ddia\, 3-2 beam. The agreement between the \dia\, and \ddia\, $T_{ex}$ values when incorporating the beam ratios in Equation \ref{eqn:wRat} suggests that in most cases, the \ddia\, 3-2 beam is similar in size to the emitting regions. }We expect, therefore, that a low filling factor would decrease our derived $R_D$ values by a maximum of $\sim 2$ for a few sources. \showrev{While we therefore expect that the introduced scatter should be small, it may be significant if the \ddia\, source size is correlated with other core properties.} This effect could be probed through high resolution maps of the protostellar sources in \dia\, and \ddia\, emission. 

\subsubsection{Offsets of \ddia\, emission from the dust continuum peak and abundance gradients}

Offsets between the position of peak \ddia\, emission and our pointing locations would be an additional source of scatter in our results. As discussed in \S\ref{sec:targets}, our target list is dominated by cores identified based on SCUBA 850\,\micron\, continuum emission. Through additional observations in a cross pattern around many of their targets, \citet{kirk07} showed that the peak \dia\, emission toward these cores is coincident with the continuum peak for nearly all their SCUBA sources. The targets selected based on extinction data were more likely to have brighter \dia\, emission offset from their central pointing location. Of the 13 sources in our sample not selected from SCUBA data, only three show substantially stronger \dia\, emission $\sim 25$\arcsec\, offset to our pointing locations. We thus are confident we are probing the peak of the \dia\, emission for the large majority of our sources. 

Some studies have shown that the emission of deuterated species can be significantly offset from associated submillimeter continuum emission, or even from their non-deuterated counterparts \citep{pagani05,friesen10b}. Given the beam size of the observations presented here (30\arcsec\, and 41\arcsec), the \ddia\, emission would have to be substantially offset from the \dia\, emission to be problematic for this analysis. Based on 850\,\micron\, continuum maps, the median effective radius of our targets is 28\arcsec. Since \ddia\, emission should remain associated with the core, even if offset from the central peak as measured by submillimeter continuum or \dia\, emission, we expect scatter due to emission offsets to be small. Small maps of \dia\, and \ddia\, toward the cores would be necessary to quantify this effect. 

\begin{deluxetable*}{crrrccc}
\tablecolumns{7}
\tabletypesize{\scriptsize}
\tablecaption{Deuterium fractionation and CO depletion in the Perseus cores \label{tab:deutFrac}}
\tablehead{
\colhead{Source} & \colhead{$N(\mbox{\dia})$} & 
\colhead{$N(\mbox{\ddia})$} & 
\colhead{$R_D$} & \colhead{$N(\mbox{H}_2$)} & 
\colhead{$N(\mbox{C$^{18}$O})$} & \colhead{$f_D$} \\
\colhead{} & \colhead{$\times 10^{11}$\,cm$^{-2}$} &
\colhead{$\times 10^{11}$\,cm$^{-2}$} &
\colhead{} & 
\colhead{$\times 10^{22}$\,cm$^{-2}$} & 
\colhead{$\times 10^{14}$\,cm$^{-2}$} & \colhead{} }
\startdata 
\multicolumn{7}{c}{Detected in both \ddia\, 2-1 and 3-2} \\
\cline {1-7}
  6 &  97  &  19.2 (1.4) & 0.198 (0.042) &   1.7  &  20.9  &  1.41  \\
 21 &  81  &   6.2 (0.3) & 0.077 (0.016) &   1.6  &  29.9  &  0.92  \\
 23 &  56  &   7.8 (1.5) & 0.140 (0.038) &   1.3  &  36.1  &  0.63  \\
 25 & 125  &  23.8 (0.7) & 0.190 (0.038) &   4.6  &  37.3  &  2.10  \\
 26 & 105  &   8.2 (0.3) & 0.078 (0.016) &   2.4  &  17.1  &  2.38  \\
 27 &  42  &   4.9 (0.8) & 0.116 (0.029) &   1.1  &  25.9  &  0.73  \\
 36 &  59  &   2.6 (1.0) & 0.044 (0.019) &   0.3  &   8.4  &  0.66  \\
 71 & 355  &  33.9 (1.1) & 0.096 (0.019) &   6.7  &  24.6  &  4.65  \\
 72 & 248  &  17.1 (0.5) & 0.069 (0.014) &   6.5  &  41.6  &  2.64  \\
 73 & 343  &  31.7 (1.2) & 0.092 (0.019) &   4.0  &  32.7  &  2.09  \\
 74 & 128  &  12.1 (1.3) & 0.095 (0.022) &   2.0  &  14.6  &  2.28  \\
 75 & 203  &   9.0 (1.9) & 0.044 (0.013) &   1.8  &  22.2  &  1.35  \\
 77 & 115  &   7.4 (1.5) & 0.064 (0.018) &   1.3  &  18.2  &  1.23  \\
 86 & 171  &  14.2 (0.6) & 0.083 (0.017) &   6.3  &  22.7  &  4.72  \\
 90 &  66  &  10.7 (0.5) & 0.163 (0.033) &   3.0  &  13.7  &  3.74  \\
 92 & 193  &  11.3 (1.5) & 0.059 (0.014) &   0.8  &   7.4  &  1.89  \\
 95 & 126  &   9.5 (0.8) & 0.075 (0.016) &   1.7  &  14.9  &  1.98  \\
 96 & 135  &  13.8 (1.2) & 0.102 (0.022) &   1.3  &  11.4  &  1.96  \\
 97 & 116  &   7.0 (0.6) & 0.060 (0.013) &   1.7  &  12.7  &  2.23  \\
 99 &  48  &   5.3 (0.7) & 0.111 (0.026) &   1.3  &  11.3  &  1.98  \\
100 &  80  &   5.2 (0.5) & 0.065 (0.014) &   1.3  &  31.1  &  0.71  \\
102 & 229  &  14.6 (0.7) & 0.064 (0.013) &   2.2  &  35.2  &  1.06  \\
103 & 104  &  20.3 (0.6) & 0.196 (0.039) &  12.5  &  11.3  & 18.89  \\
104 & 142  &   4.6 (0.3) & 0.032 (0.007) &   4.1  &  38.8  &  1.81  \\
106 & 147  &  22.9 (0.8) & 0.156 (0.032) &  18.4  &  38.6  &  8.08  \\
107 & 209  &  24.4 (0.5) & 0.117 (0.024) &   3.6  &  27.0  &  2.28  \\
109 &  77  &   9.2 (0.5) & 0.120 (0.025) &   1.5  &  36.3  &  0.72  \\
110 & 150  &  10.8 (0.3) & 0.072 (0.015) &   3.0  &  45.6  &  1.11  \\
111 & 233  &   6.3 (0.3) & 0.027 (0.006) &   1.6  &  16.5  &  1.68  \\
112 & 233  &   7.7 (0.3) & 0.033 (0.007) &   7.9  &  56.4  &  2.39  \\
113 & 176  &  14.3 (0.5) & 0.081 (0.016) &   3.3  &  74.9  &  0.75  \\
115 & 128  &   5.4 (0.5) & 0.042 (0.009) &   0.6  &  19.0  &  0.57  \\
116 & 113  &   7.6 (0.3) & 0.067 (0.014) &   2.8  &  48.7  &  0.98  \\
118 & 161  &   6.0 (0.5) & 0.037 (0.008) &   4.5  &  39.7  &  1.91  \\
121 & 100  &   5.2 (0.6) & 0.052 (0.012) &   1.3  &  15.1  &  1.49  \\
122 & 210  &   7.3 (0.7) & 0.035 (0.008) &   2.3  &  14.7  &  2.68  \\
123 & 208  &   8.8 (1.0) & 0.042 (0.010) &   2.3  &  21.4  &  1.80  \\
124 & 124  &   6.3 (0.8) & 0.051 (0.012) &   1.4  &  24.0  &  1.01  \\
134 & 158  &   7.0 (0.6) & 0.044 (0.010) &   2.3  &  10.1  &  3.81  \\
135 & 206  &   9.0 (1.1) & 0.044 (0.010) &   2.0  &   6.5  &  5.26  \\
136 &  89  &   3.3 (1.1) & 0.038 (0.014) &   1.1  &   7.3  &  2.57  \\
146 & 132  &  13.4 (1.1) & 0.101 (0.022) &   3.0  &  22.1  &  2.29  \\
148 & 169  &  22.8 (1.1) & 0.135 (0.028) &   4.0  &   9.5  &  7.07  \\
149 & 332  &  25.1 (0.7) & 0.076 (0.015) &  10.5  &  42.6  &  4.20  \\
150 & 253  &  15.8 (1.8) & 0.063 (0.014) &   2.5  &  17.6  &  2.41  \\
152 & 313  &  10.4 (1.5) & 0.033 (0.008) &   4.3  &  24.5  &  2.95  \\
\cline {1-7}
\multicolumn{7}{c}{Detected in \ddia\, 2-1 only} \\
\cline {1-7}
 36 &  59  &   1.7 (0.3) & 0.029 (0.008) &   0.3  &   8.4  &  0.66  \\
147 &  83  &   5.0 (0.4) & 0.060 (0.013) &   0.5  &  22.3  &  0.37  \\
\cline {1-7}
\multicolumn{7}{c}{Detected in \ddia\, 3-2 only} \\
\cline {1-7}
  3 & 145  &   4.1 (0.7) & 0.029 (0.008) &   0.3  &  13.1  &  0.43  \\
  4 & 145  &   4.1 (0.5) & 0.028 (0.007) &   1.6  &  23.3  &  1.19  \\
 15 &  55  &   4.4 (0.9) & 0.080 (0.023) &   1.2  &  23.3  &  0.87  \\
 19 &  66  &   1.6 (0.4) & 0.024 (0.008) &   1.5  &  19.1  &  1.35  \\
143 & 114  &   5.0 (1.4) & 0.044 (0.015) &   1.0  &   5.3  &  3.25  \\
\cline {1-7}
\multicolumn{7}{c}{Non-detection upper limits} \\
\cline {1-7}
  5 &  33 & $\le$   1.5 & $\le$ 0.030 &   0.1  &  18.9  &  0.13  \\
 13 &  42 & $\le$   0.8 & $\le$ 0.018 &  \nodata  &  \nodata  &  \nodata  \\
 79 & 126 & $\le$   6.2 & $\le$ 0.035 &   1.3  &  24.2  &  0.89  \\
 84 &  51 & $\le$   1.5 & $\le$ 0.029 &   0.9  &  17.0  &  0.90  \\
 85 & 190 & $\le$  14.0 & $\le$ 0.074 &   1.0  &  17.1  &  1.01  \\
 94 &  49 & $\le$   0.9 & $\le$ 0.019 &   0.8  &   9.6  &  1.44  \\
125 &  98 & $\le$   2.1 & $\le$ 0.018 &   2.2  &  17.5  &  2.12  \\
126 &  54 & $\le$   1.3 & $\le$ 0.020 &   0.8  &   6.5  &  2.05  \\
128 & 117 & $\le$   1.8 & $\le$ 0.018 &   1.0  &  10.7  &  1.63  \\
130 &  27 & $\le$   2.1 & $\le$ 0.107 &   0.2  &   2.3  &  1.31  \\
131 &  29 & $\le$   1.4 & $\le$ 0.045 &   \nodata  &   2.6  & \nodata  \\
133 & 128 & $\le$   3.7 & $\le$ 0.025 &   1.6  &   7.2  &  3.80  \\
\enddata
\tablecomments{Uncertainties $N(\mbox{\dia})$, $N(\mbox{H}_2)$, and $N(\mbox{CO})$ are discussed in the text.}
\end{deluxetable*}

\subsection{Comparison with chemistry models}
\label{sec:chem}

Despite the lack of clear trends in the deuterium fractionation of the Perseus cores with many parameters that may trace an evolutionary sequence, the data nevertheless inhabit a parameter space that needs to be accounted for in models that combine physical and chemical evolution of low-mass star forming cores. In general, our results are in broad agreement with published steady-state and dynamic chemical models of core evolution. The deuterium fractionation values for the Perseus cores are bracketed by the slow and fast collapse models of \citet{aikawa05}, who show that cores that collapse quickly will show less deuterium fractionation in \dia\, compared with cores that collapse more slowly. The radial velocity profiles predicted for the fast collapse models are large, however, compared with typical observational results. While large infall speeds have been found for the protostellar sources NGC 1333 IRAS 4A and 4B \citep[0.7\,\kms\, and 0.5\,\kms, respectively;][]{difran01}, more typical infall speeds are $\sim 0.05$\,\kms\, to 0.1\,\kms\, \citep{mardones98}. Steady-state chemistry core models can reproduce the range of $R_D$ values found in Perseus given variations in the model parameters \citep{caselli08}. The dominant parameters that affect the deuterium fractionation at the low temperatures ($T \lesssim 17$\,K) characteristic of our targets in Perseus are the CO depletion, the core density, the fractional abundance of dust grains with respect to H$_2$, the cosmic-ray ionization rate, and the ratio of ortho-H$_2$ to para-H$_2$ in the gas \citep{pagani92,gerlich02,walmsley04,dalgarno, caselli08,roueff05,roberts00}. 

The amount of CO depletion in cold, dense gas strongly impacts the level of deuterium fractionation that can be reached \citep{dalgarno, caselli08}. Using a simple chemical model, \citet{keto08} calculate the expected steady-state CO abundance as a function of density and temperature in starless cores. Given their average densities and temperatures, we expect CO depletion factors $f_D \sim 6-8$ for the Perseus cores. In \S\ref{sec:co}, however, we find that the CO depletion factors for most of the Perseus cores are low (mean $f_D = 3.8$), with all but three cores having depletion factors $f_D < 5$. In fact, we find an over-abundance of CO relative to H$_2$ toward twelve cores, compared with the canonical abundance ratio of $[\mbox{C$^{18}$O}] \ [\mbox{H$_2$}] = 1.7 \times 10^{-7}$ \citep{frerking82}. \showrev{The CO depletion measurement is dependent on the conversion between observed submillimeter continuum emission and H$_2$ column density. As described in \S\ref{sec:co}, several parameters in this conversion are not known to high precision.}

It is also possible that the canonical abundance ratio of CO to H$_2$ may vary between nearby molecular clouds. For example, \citet{tafalla_santiago} find a best fit C$^{18}$O abundance relative to H$_2$, $X(\mbox{C$^{18}$O}) = 0.7 \times 10^{-7}$ toward the starless core L1521E in Taurus, which they argue is undepleted in CO. \citet{lacy94} observe directly the vibrational absorption of both H$_2$ and CO toward two infrared sources in Orion, and find $[\mbox{CO}]/[\mbox{H$_2$}] = 2.7 \times 10^{-4}$, resulting in an abundance ratio for C$^{18}$O of $\sim 4.8 \times 10^{-7}$ assuming $[\mbox{CO}]/[\mbox{C$^{18}$O}] = 560 \pm 25$ \citep{wilson94}. Using the \citeauthor{lacy94} result then gives $f_D > 1$ for all but three cores, and increases the mean C$^{18}$O value to $f_D = 8$, in agreement with \citet{keto08}. 

We next compare the Perseus CO depletion results to the recent work by \citet{christie12}, who have performed the first large-scale, systematic study of the CO depletion toward star-forming cores in the Taurus, Serpens, Ophiuchus, and Orion molecular clouds as part of the JCMT Gould Belt Legacy Survey \citep{ward-thompson07}. They found that while cores with high levels of depletion are found within each cloud, there is wide variation in the typical CO depletion found across each cloud. For example, Taurus is characterized by relatively high depletion values (mean $f_D = 25$ for starless sources), while smaller $f_d$ values are seen in Ophiuchus (mean $f_D = 7$ for starless sources, and $f_D = 4$ for protostellar sources), similar to our findings in Perseus. The authors suggest the low $f_D$ values in Ophiuchus are related to a possible anomalous grain size distribution in the cloud \citep{carrasco73}. Depletion of CO can be suppressed by the presence of small grains, because any CO accreted onto the grain will quickly desorb by thermal evaporation \citep{leger85}. Intriguingly, the presence of anomalous dust emission has also been noted in Perseus, where microwave observations show evidence of a population of small, spinning dust grains in the dense molecular gas on degree scales \citep{watson05,ade11}. \showrev{This spinning dust emission is associated, however, with the lower density envelopes of the molecular clouds themselves, so it is unclear how the small grains responsible for the microwave emission would impact the depletion of CO toward the cores themselves.}

Increasing central density should also lead to greater deuterium fractionation \citep{roueff05,aikawa05,caselli08}. In agreement, in \S \ref{sec:trends}, we find a significant correlation between increasing $R_D$ and central core density for starless cores. The deeply embedded sources (those not detected in Spitzer catalogs of protostellar sources in Perseus) appear to fall along the starless core trend. Protostellar cores, in contrast, show a scatter of $R_D$ values even at high central densities. Comparing Figures \ref{fig:rdTrends} and \ref{fig:lbol}, we see that the four protostellar sources showing the greatest levels of deuterium fractionation have high central densities (based on BE modeling), and are also very young based on their low $L_{bol}$ and $T_{bol}$, while retaining significant material in their envelopes (envelope masses $M_{env} \gtrsim M_\odot$). Figure \ref{fig:lbol} also shows that no protostellar cores have $R_D > 0.05$ at $L_{bol}/F_{1.1\,\mbox{mm}} > 3$, with only two sources showing high $R_D$ values at $L_{bol}/F_{1.1\,\mbox{mm}} > 1$. Multiple protostellar cores with $L_{bol}/F_{1.1\,\mbox{mm}} < 1$ also have low $R_D$, however. The scatter in the $R_D$ values against $L_{bol}$ and $T_{bol}$ may be the result of differences between individual cores in how the core evolved from starless to protostellar. For example, the maximum $R_D$ reached at the starless core stage could vary between cores depending on the timescale over which the core condensed and collapsed. Nonetheless, the data are consistent with the general evolutionary scenario where the deuterium fractionation in starless cores increases with central density, and decreases with time after a protostar has formed and is detected in infrared emission. 

The other parameters are less certain for the Perseus cores, and may be the cause of the large scatter in $R_D$ in the cores \emph{if} they vary across the Perseus molecular cloud. \citet{caselli08} show that variations in the fractional abundance of dust grains will impact the deuterium fractionation, where a decrease in the minimum radius of dust grains in a \citet{mathis77} size distribution leads to an decrease in the deuterium fractionation. We have noted above the possibility that an anomalous population of small dust grains in Perseus is the cause of the low CO depletion. The deuterium fractionation will also decrease with increased cosmic-ray ionization rate, as the increased number of electrons will dissociate \hd\, before the deuterium can be transferred to other species, such as \dia\, \citep{roberts00}. At an angular resolution of several arcminutes, the dust temperatures are greater by several Kelvin in the east \citep{schnee05}, which \citet{ridge06} argue is due to heating from an H{\small{II}} region driven by the nearby early-B star HD 278942 (although note that both IC348 and NGC 1333 show localized higher temperatures). On core scales, the gas temperatures determined through analysis of the \amm\, (1,1) and (2,2) lines are significantly elevated only toward NGC 1333. While the elevated dust temperatures on large scales thus do not appear to affect the intrinsic core temperatures on small scales, they provide evidence of variation in the interstellar radiation field (ISRF) across the Perseus molecular cloud. A significantly stronger ISRF would increase the electron abundance, however, and consequently decrease the abundance of molecular ions (such as \dia) through dissociative recombination reactions. As discussed in \S\ref{sec:rd}, we do not see a systematic variation in \dia\, abundances in the Perseus cores across the molecular cloud. 

Finally, high ratios of ortho-H$_2$ to para-H$_2$ in the gas will also lower the abundance of deuterated species by decreasing the \hd\, abundance \citep{pagani92,gerlich02,walmsley04,flower06,caselli08,pagani09,pagani11}. In contrast to CO, the timescales for some of the chemical reactions involved in the deuterium fractionation process are longer than the dynamical timescale. In particular, H$_2$ forms with an ortho- to para- ratio of 3:1, and does not reach the steady-state ratio of $2.7 \times 10^{-3}$ (assuming constant $T=10$\,K and $n=10^4$\,\cc) for $t \sim 10^7$\,yr  \citep{flower06}. In regions like Perseus, the ortho- to para-H$_2$ ratio, and consequently the deuterium fractionation of species like \dia, will depend on the condensation timescale and thermal history of the individual cores. Lower deuterium fractionation values may therefore indicate that the cores in Perseus are young ($t < 10^6$\,yr), in agreement with age estimates by \citet{jorgensen08} based on infrared and continuum analysis. 

\section{Summary}
\label{sec:summary}

We have performed a systematic study of the deuterium fractionation of 64 dense cores within the Perseus molecular cloud through analysis of \ddia\, observations and published \dia\, results, detecting 54 sources in \ddia\, 2-1, \ddia\, 3-2, or both. In the detected sources, we find a mean \dia\, deuterium fraction, $R_D = N(\mbox{\ddia})\,/\,N(\mbox{\dia})$, of 0.08. We find no significant difference in the \dia\, deuterium fraction between starless and protostellar cores, and no difference based on core location in clustered or isolated environments, although isolated cores were least likely to be detected in \ddia\, emission. 

Kinematically, we find that \dia\, and \ddia\, exhibit very similar behaviour in the Perseus cores.  Both have similar, very small, velocity differences with respect to \amm\ (typically less than $\sim 0.07$\,\kms, and both also have similar, slightly larger, velocity differences with respect to \co\ (typically less than $\sim 0.19$\,\kms); for a temperature of 15~K, the sound speed is 0.23\,\kms.  The small centroid velocity differences between \ddia, \dia, and \amm\ suggest that these nitrogen-bearing species all occupy a similar volume of each dense core, despite the differences in critical densities.

We compare the deuterium fractionation levels with parameters linked to advanced core evolution, and find a few significant correlations. First, we find that greater \dia\, deuterium fraction correlates with increased H$_2$ column density and increased central core density, when considering the entire core sample. The trend with $N(\mbox{H}_2)$ is stronger when limited to only protostellar cores, while the trend with $\log(n_c)$ is stronger when limited to only starless cores. We additionally find that the \dia\, deuterium fraction decreases with increased $T_{bol}$ for protostellar sources. Significant correlations were not found between $R_D$ and other parameters, including \dia\, line width, kinetic gas temperature, or protostellar bolometric luminosity. The data are nevertheless in agreement with the general expectation that the \dia\, deuterium fraction will decrease toward protostellar cores as the embedded protostar warms its immediate environment

We show that the Perseus cores are characterized by low CO depletion values relative to previous studies of star forming cores, with a mean $f_D$ of 3.1. The CO depletion values are similar, however, to those found for cores in the Ophiuchus molecular cloud. We suggest that the low average CO depletion factor is the dominant mechanism that constrains the average deuterium fractionation in the Perseus cores to small values. While current equilibrium and dynamic chemical models are able to reproduce the range of deuterium fractionation values we find in Perseus, reproducing the scatter across the cores requires variation in parameters such as the ionization fraction or the ortho- to para-H$_2$ ratio across the cloud, or a range in core evolution timescales. 

\acknowledgments

The authors thank the anonymous referee, whose thoughtful comments and suggestions greatly improved the manuscript. The authors also thank P. Caselli and D. Johnstone for helpful discussion and comments. RF is a Dunlap Fellow at the Dunlap Institute for Astronomy \& Astrophysics, University of Toronto. The Dunlap Institute is funded through an endowment established by the David Dunlap family and the University of Toronto. HK acknowledges support from the Smithsonian Scholarly Studies Program and a Natural Sciences and Engineering Council of Canada Postdoctoral Fellowship. The SMT and Kitt Peak 12 Meter are operated by the Arizona Radio Observatory (ARO), Steward Observatory, University of Arizona.

{\it Facilities:} \facility{ARO:12m}, \facility{HHT}


\end{document}